\input harvmac   
\noblackbox   

\input labeldefs.tmp


\let\includefigures=\iftrue
\let\useblackboard=\iftrue
\newfam\black

\includefigures
\message{If you do not have epsf.tex (to include figures),}
\message{change the option at the top of the tex file.}
\input epsf
\def\figin{\epsfcheck\figin}\def\figins{\epsfcheck\figins}

\def\epsfcheck{\ifx\epsfbox\UnDeFiNeD
\message{(NO epsf.tex, FIGURES WILL BE IGNORED)}
\gdef\figin##1{\vskip2in}\gdef\figins##1{\hskip.5in}
\else\message{(FIGURES WILL BE INCLUDED)}%
\gdef\figin##1{##1}\gdef\figins##1{##1}\fi}

\def\DefWarn#1{}
\def\figinsert{\goodbreak\midinsert}

\def\ifig#1#2#3{\DefWarn#1\xdef#1{fig.~\the\figno}
\writedef{#1\leftbracket fig.\noexpand~\the\figno}%
\figinsert\figin{\centerline{#3}}\medskip\centerline{\vbox{
\baselineskip12pt\advance\hsize by -1truein
\noindent\footnotefont{\bf Fig.~\the\figno:} #2}}
\bigskip\endinsert\global\advance\figno by1}
\else
\def\ifig#1#2#3{\xdef#1{fig.~\the\figno}
\writedef{#1\leftbracket fig.\noexpand~\the\figno}%
\global\advance\figno by1}
\fi

\def\p{\partial}

\def\CN{{\cal N}}

\def\CS{{\cal S}}

\def\CS{{\cal S }}

\def\G{\Gamma}    


\font\manual=manfnt     
\def\dbend{\lower3.5pt\hbox{\manual\char127}}

\def\IZ{\relax\ifmmode\mathchoice    
{\hbox{\cmss Z\kern-.4em Z}}{\hbox{\cmss Z\kern-.4em Z}}    
{\lower.9pt\hbox{\cmsss Z\kern-.4em Z}} {\lower1.2pt\hbox{\cmsss    
Z\kern-.4em Z}}\else{\cmss Z\kern-.4em Z}\fi}    
\def\half {{1\over 2}}

\def\p{\partial}    
    
\def\bar{\overline}    
\def\CS{{\cal S}}    
\def\CN{{\cal N}}    
    
\def\rt2{\sqrt{2}}    
\def\irt2{{1\over\sqrt{2}}}

\def\t{\tilde}    
\def\ndt{\noindent}

\def\a{\alpha}

\font\cmss=cmss10    
\font\cmsss=cmss10 at 7pt    
\def\IL{\relax{\rm I\kern-.18em L}}    
\def\IH{\relax{\rm I\kern-.18em H}}    
\def\IR{\relax{\rm I\kern-.18em R}}    
\def\inbar{\vrule height1.5ex width.4pt depth0pt}    
\def\IC{\relax\hbox{$\inbar\kern-.3em{\rm C}$}}    
\def\rlx{\relax\leavevmode}    
\def\ZZ{\rlx\leavevmode\ifmmode\mathchoice{\hbox{\cmss Z\kern-.4em Z}}    
 {\hbox{\cmss Z\kern-.4em Z}}{\lower.9pt\hbox{\cmsss Z\kern-.36em Z}}    
 {\lower1.2pt\hbox{\cmsss Z\kern-.36em Z}}\else{\cmss Z\kern-.4em    
 Z}\fi}     
\def\IZ{\relax\ifmmode\mathchoice    
{\hbox{\cmss Z\kern-.4em Z}}{\hbox{\cmss Z\kern-.4em Z}}    
{\lower.9pt\hbox{\cmsss Z\kern-.4em Z}}    
{\lower1.2pt\hbox{\cmsss Z\kern-.4em Z}}\else{\cmss Z\kern-.4em    
Z}\fi}    
    

\def\G{\Gamma}

\font\manual=manfnt     
\def\dbend{\lower3.5pt\hbox{\manual\char127}}

\def\IZ{\relax\ifmmode\mathchoice    
{\hbox{\cmss Z\kern-.4em Z}}{\hbox{\cmss Z\kern-.4em Z}}    
{\lower.9pt\hbox{\cmsss Z\kern-.4em Z}} {\lower1.2pt\hbox{\cmsss    
Z\kern-.4em Z}}\else{\cmss Z\kern-.4em Z}\fi}    
\def\half {{1\over 2}}

\def\bar{\overline}    

\def\rt2{\sqrt{2}}    
\def\irt2{{1\over\sqrt{2}}}

\def\t{\tilde}    
\def\T{\widetilde}
\def\ndt{\noindent}

\def\bra#1{{\langle}#1|}
\def\ket#1{|#1\rangle}
\def\bbra#1{{\langle\langle}#1|}
\def\kket#1{|#1\rangle\rangle}

%


\def\doublefig#1#2#3#4{\DefWarn#1\xdef#1{fig.~\the\figno}
\writedef{#1\leftbracket fig.\noexpand~\the\figno}%
\figinsert\figin{\centerline{#3\hskip1.0cm#4}}\medskip\centerline{\vbox{
\baselineskip12pt\advance\hsize by -1truein
\noindent\footnotefont{\bf Fig.~\the\figno:} #2}}
\bigskip\endinsert\global\advance\figno by1}


\writedefs


\lref\McGreevyEP{
  J.~McGreevy, J.~Teschner and H.~Verlinde,
  ``Classical and quantum D-branes in 2D string theory,''
  JHEP {\bf 0401}, 039 (2004)
  [arXiv:hep-th/0305194].
}

\lref\HananyIE{
  A.~Hanany and E.~Witten,
  ``Type IIB superstrings, BPS monopoles, and three-dimensional gauge dynamics,''
  Nucl.\ Phys.\ B {\bf 492}, 152 (1997)
  [arXiv:hep-th/9611230].
}

\lref\PolchinskiMT{
  J.~Polchinski,
  ``Dirichlet-Branes and Ramond-Ramond Charges,''
  Phys.\ Rev.\ Lett.\  {\bf 75}, 4724 (1995)
  [arXiv:hep-th/9510017].
}

\lref\EguchiTC{
T.~Eguchi and Y.~Sugawara,
``Modular invariance in superstring on Calabi-Yau n-fold with A-D-E
singularity,''
Nucl.\ Phys.\ B {\bf 577}, 3 (2000)
[arXiv:hep-th/0002100].
}

\lref\sameer{
S.~Murthy,
``Notes on non-critical superstrings in various dimensions,''
JHEP {\bf 0311}, 056 (2003)
[arXiv:hep-th/0305197].
}

\lref\KlebanovYA{
I.~R.~Klebanov and J.~M.~Maldacena,
``Superconformal gauge theories and non-critical superstrings,''
arXiv:hep-th/0409133.
}

\lref\KazamaQP{
  Y.~Kazama and H.~Suzuki,
  ``New N=2 Superconformal Field Theories And Superstring Compactification,''
  Nucl.\ Phys.\ B {\bf 321}, 232 (1989).
}

\lref\ElitzurPQ{
S.~Elitzur, A.~Giveon, D.~Kutasov, E.~Rabinovici and G.~Sarkissian,
``D-branes in the background of NS fivebranes,''
JHEP {\bf 0008}, 046 (2000)
[arXiv:hep-th/0005052].
}

\lref\NamHU{
S.~k.~Nam,
``Superconformal And Super Kac-Moody Invariant Quantum Field Theories In
Two-Dimensions,''
Phys.\ Lett.\ B {\bf 187}, 340 (1987).
}

\lref\KiritsisRV{
E.~Kiritsis,
``Character Formulae And The Structure Of The Representations Of The N=1, N=2
Superconformal Algebras,''
Int.\ J.\ Mod.\ Phys.\ A {\bf 3}, 1871 (1988).
}

\lref\WittenYR{
E.~Witten,
``On string theory and black holes,''
Phys.\ Rev.\ D {\bf 44}, 314 (1991).
}

\lref\DijkgraafBA{
R.~Dijkgraaf, H.~Verlinde and E.~Verlinde,
``String propagation in a black hole geometry,''
Nucl.\ Phys.\ B {\bf 371}, 269 (1992).
}

\lref\HoriAX{
K.~Hori and A.~Kapustin,
``Duality of the fermionic 2d black hole and N = 2 Liouville theory as  mirror
symmetry,''
JHEP {\bf 0108}, 045 (2001)
[arXiv:hep-th/0104202].
}

\lref\AhnTT{
  C.~Ahn, M.~Stanishkov and M.~Yamamoto,
  ``One-point functions of N = 2 super-Liouville theory with boundary,''
  Nucl.\ Phys.\ B {\bf 683}, 177 (2004)
  [arXiv:hep-th/0311169].
}

\lref\KutasovUA{
D.~Kutasov and N.~Seiberg,
``Noncritical Superstrings,''
Phys.\ Lett.\ B {\bf 251}, 67 (1990).
}

\lref\EguchiYI{
T.~Eguchi and Y.~Sugawara,
``SL(2,R)/U(1) supercoset and elliptic genera of non-compact Calabi-Yau
manifolds,''
JHEP {\bf 0405}, 014 (2004)
[arXiv:hep-th/0403193].
}

\lref\FotopoulosUT{
A.~Fotopoulos, V.~Niarchos and N.~Prezas,
``D-branes and extended characters in SL(2,R)/U(1),''
Nucl.\ Phys.\ B {\bf 710}, 309 (2005)
[arXiv:hep-th/0406017].
}

\lref\HarmarkSF{
  T.~Harmark, V.~Niarchos and N.~A.~Obers,
  ``Stable non-supersymmetric vacua in the moduli space of non-critical
  superstrings,''
  Nucl.\ Phys.\  B {\bf 759}, 20 (2006)
  [arXiv:hep-th/0605192].
}

\lref\GaberdielJR{
M.~R.~Gaberdiel,
``Lectures on non-BPS Dirichlet branes,''
Class.\ Quant.\ Grav.\  {\bf 17}, 3483 (2000)
[arXiv:hep-th/0005029].
}

\lref\FotopoulosUT{
  A.~Fotopoulos, V.~Niarchos and N.~Prezas,
  ``D-branes and extended characters in SL(2,R)/U(1),''
  Nucl.\ Phys.\ B {\bf 710}, 309 (2005)
  [arXiv:hep-th/0406017].
}

\lref\FZZ{
V.~A.~Fateev, A.~B.~Zamolodchikov and Al.~B.~Zamolodchikov, unpublished.
}

\lref\AharonyUB{
O.~Aharony, M.~Berkooz, D.~Kutasov and N.~Seiberg,
``Linear dilatons, NS5-branes and holography,''
JHEP {\bf 9810}, 004 (1998)
[arXiv:hep-th/9808149].
}

\lref\AharonyXN{
  O.~Aharony, A.~Giveon and D.~Kutasov,
  ``LSZ in LST,''
  Nucl.\ Phys.\  B {\bf 691}, 3 (2004)
  [arXiv:hep-th/0404016].
}

\lref\RecknagelSB{
  A.~Recknagel and V.~Schomerus,
  ``D-branes in Gepner models,''
  Nucl.\ Phys.\ B {\bf 531}, 185 (1998)
  [arXiv:hep-th/9712186].
}

\lref\GutperleHB{
  M.~Gutperle and Y.~Satoh,
  ``D-branes in Gepner models and supersymmetry,''
  Nucl.\ Phys.\ B {\bf 543}, 73 (1999)
  [arXiv:hep-th/9808080].
}

\lref\WittenSC{
  E.~Witten,
  ``Solutions of four-dimensional field theories via M-theory,''
  Nucl.\ Phys.\ B {\bf 500}, 3 (1997)
  [arXiv:hep-th/9703166].
}

\lref\ElitzurPQ{
S.~Elitzur, A.~Giveon, D.~Kutasov, E.~Rabinovici and G.~Sarkissian,
``D-branes in the background of NS fivebranes,''
JHEP {\bf 0008}, 046 (2000)
[arXiv:hep-th/0005052].
}

\lref\FateevIK{
V.~Fateev, A.~B.~Zamolodchikov and A.~B.~Zamolodchikov,
``Boundary Liouville field theory. I: Boundary state and boundary  two-point
function,''
arXiv:hep-th/0001012.
}

\lref\TeschnerMD{
J.~Teschner,
``Remarks on Liouville theory with boundary,''
arXiv:hep-th/0009138.
}

\lref\ZamolodchikovAH{
A.~B.~Zamolodchikov and A.~B.~Zamolodchikov,
``Liouville field theory on a pseudosphere,''
arXiv:hep-th/0101152.
}

\lref\RibaultSS{
S.~Ribault and V.~Schomerus,
``Branes in the 2-D black hole,''
JHEP {\bf 0402}, 019 (2004)
[arXiv:hep-th/0310024].
}

\lref\AhnTT{
C.~Ahn, M.~Stanishkov and M.~Yamamoto,
``One-point functions of N = 2 super-Liouville theory with boundary,''
Nucl.\ Phys.\ B {\bf 683}, 177 (2004)
[arXiv:hep-th/0311169].
}

\lref\IsraelJT{
D.~Israel, A.~Pakman and J.~Troost,
``D-branes in N = 2 Liouville theory and its mirror,''
arXiv:hep-th/0405259.
}

\lref\AhnQB{
C.~Ahn, M.~Stanishkov and M.~Yamamoto,
``ZZ-branes of N = 2 super-Liouville theory,''
JHEP {\bf 0407}, 057 (2004)
[arXiv:hep-th/0405274].
}

\lref\hoso{
K.~Hosomichi,
``N = 2 Liouville theory with boundary,''
arXiv:hep-th/0408172.
}

\lref\IsraelFN{
D.~Israel, A.~Pakman and J.~Troost,
``D-branes in little string theory,''
arXiv:hep-th/0502073.
}

\lref\EguchiIK{
  T.~Eguchi and Y.~Sugawara,
  ``Modular bootstrap for boundary N = 2 Liouville theory,''
  JHEP {\bf 0401}, 025 (2004)
  [arXiv:hep-th/0311141].
}

\lref\BilalUH{
A.~Bilal and J.~L.~Gervais,
``New Critical Dimensions For String Theories,''
Nucl.\ Phys.\ B {\bf 284}, 397 (1987).
}

\lref\BilalIA{
A.~Bilal and J.~L.~Gervais,
``Modular Invariance For Closed Strings At The New Critical Dimensions,''
Phys.\ Lett.\ B {\bf 187}, 39 (1987).
}

\lref\BarsSR{
I.~Bars and K.~Sfetsos,
``Conformally exact metric and dilaton in string theory on curved
space-time,''
Phys.\ Rev.\ D {\bf 46}, 4510 (1992)
[arXiv:hep-th/9206006].
}

\lref\TseytlinMY{
A.~A.~Tseytlin,
``Conformal sigma models corresponding to gauged Wess-Zumino-Witten
theories,''
Nucl.\ Phys.\ B {\bf 411}, 509 (1994)
[arXiv:hep-th/9302083].
}

\lref\Teschner{
  J.~Teschner,
  ``On boundary perturbations in Liouville theory and brane dynamics in
  noncritical string theories,''
  JHEP {\bf 0404}, 023 (2004)
  [arXiv:hep-th/0308140].
}

\lref\GiveonSR{
  A.~Giveon and D.~Kutasov,
  ``Brane dynamics and gauge theory,''
  Rev.\ Mod.\ Phys.\  {\bf 71}, 983 (1999)
  [arXiv:hep-th/9802067].
}

\lref\FotopoulosCN{
  A.~Fotopoulos, V.~Niarchos and N.~Prezas,
  ``D-branes and SQCD in non-critical superstring theory,''
  JHEP {\bf 0510}, 081 (2005)
  [arXiv:hep-th/0504010].
}

\lref\polchinski{J.~Polchinski, ``String theory, Vol. I, II'', Cambridge University Press (1998).}

\lref\farkas{H.~M.~Farkas and I.~Kra, ``Theta constants, Riemann surfaces and the modular group'', Graduate studies in mathematics, Vol.37. Amer.\ Math.\ Soc.\ }

\lref\tatatheta{D.~Mumford, ``Tata Lectures on Theta''}

\lref\MizoguchiKK{
  S.~Mizoguchi,
 ``Modular invariant critical superstrings on four-dimensional Minkowski
  space x two-dimensional black hole,''
  JHEP {\bf 0004}, 014 (2000)
  [arXiv:hep-th/0003053].
}

\lref\absteg{
M.~Abramowitz and I.~Stegun, ``Handbook of Mathematical Functions''.
}

\lref\Ashok{
  S.~K.~Ashok, S.~Murthy and J.~Troost,
  ``D-branes in non-critical superstrings and minimal super Yang-Mills in
  various dimensions,''
  Nucl.\ Phys.\  B {\bf 749}, 172 (2006)
  [arXiv:hep-th/0504079].
}

\lref\PonsotGT{
  B.~Ponsot, V.~Schomerus and J.~Teschner,
  ``Branes in the Euclidean AdS(3),''
  JHEP {\bf 0202}, 016 (2002)
  [arXiv:hep-th/0112198].
}

\lref\RibaultPQ{
  S.~Ribault,
  ``Discrete D-branes in AdS(3) and in the 2d black hole,''
  arXiv:hep-th/0512238.
}

\lref\AshokXC{
  S.~K.~Ashok, S.~Murthy and J.~Troost,
  ``Topological cigar and the c = 1 string: Open and closed,''
  JHEP {\bf 0602}, 013 (2006)
  [arXiv:hep-th/0511239].
}

\lref\MukhiZB{
  S.~Mukhi and C.~Vafa,
  ``Two-dimensional black hole as a topological coset model of c = 1 string
  theory,''
  Nucl.\ Phys.\ B {\bf 407}, 667 (1993)
  [arXiv:hep-th/9301083].
}
\lref\LeighEP{
  R.~G.~Leigh and M.~J.~Strassler,
  ``Exactly marginal operators and duality in four-dimensional N=1
  supersymmetric gauge theory,''
  Nucl.\ Phys.\ B {\bf 447}, 95 (1995)
  [arXiv:hep-th/9503121].
}

\lref\ElitzurHC{
  S.~Elitzur, A.~Giveon, D.~Kutasov, E.~Rabinovici and A.~Schwimmer,
  ``Brane dynamics and N = 1 supersymmetric gauge theory,''
  Nucl.\ Phys.\ B {\bf 505}, 202 (1997)
  [arXiv:hep-th/9704104].
}

\lref\ShifmanUA{
  M.~A.~Shifman,
  ``Nonperturbative dynamics in supersymmetric gauge theories,''
  Prog.\ Part.\ Nucl.\ Phys.\  {\bf 39}, 1 (1997)
  [arXiv:hep-th/9704114].
}

\lref\SeibergPQ{
  N.~Seiberg,
  ``Electric - magnetic duality in supersymmetric nonAbelian gauge theories,''
  Nucl.\ Phys.\ B {\bf 435}, 129 (1995)
  [arXiv:hep-th/9411149].
}

\lref\BerensteinFI{
  D.~Berenstein and M.~R.~Douglas,
  ``Seiberg duality for quiver gauge theories,''
  arXiv:hep-th/0207027.
}

\lref\GiveonPX{
A.~Giveon and D.~Kutasov,
``Little string theory in a double scaling limit,''
JHEP {\bf 9910}, 034 (1999)
[arXiv:hep-th/9909110]
\semi 
A.~Giveon, D.~Kutasov and O.~Pelc,
``Holography for non-critical superstrings,''
JHEP {\bf 9910}, 035 (1999)
[arXiv:hep-th/9907178].
}

\lref\HoriAB{
  K.~Hori, H.~Ooguri and Y.~Oz,
  ``Strong coupling dynamics of four-dimensional N = 1 gauge theories from  M
  theory fivebrane,''
  Adv.\ Theor.\ Math.\ Phys.\  {\bf 1}, 1 (1998)
  [arXiv:hep-th/9706082].
\semi
  E.~Witten,
  ``Branes and the dynamics of {QCD},''
  Nucl.\ Phys.\ B {\bf 507}, 658 (1997)
  [arXiv:hep-th/9706109].
%
\semi
  A.~Brandhuber, N.~Itzhaki, V.~Kaplunovsky, J.~Sonnenschein and S.~Yankielowicz,
  ``Comments on the M theory approach to N = 1 S{QCD} and brane dynamics,''
  Phys.\ Lett.\ B {\bf 410}, 27 (1997)
  [arXiv:hep-th/9706127].

}

\lref\StrasslerQS{
  M.~J.~Strassler,
  ``The duality cascade,''
  arXiv:hep-th/0505153.
}

\lref\BarbonZU{
  J.~L.~F.~Barbon,
  ``Rotated branes and N = 1 duality,''
  Phys.\ Lett.\ B {\bf 402}, 59 (1997)
  [arXiv:hep-th/9703051].
}

\lref\murtro{
  S.~Murthy and J.~Troost,
   ``D-branes in non-critical superstrings and duality in N = 1 gauge theories
  with flavor,''
  JHEP {\bf 0610}, 019 (2006)
  [arXiv:hep-th/0606203].
}

\lref\GaberdielNV{
  M.~R.~Gaberdiel and H.~Klemm,
  ``N = 2 superconformal boundary states for free bosons and fermions,''
  Nucl.\ Phys.\ B {\bf 693}, 281 (2004)
  [arXiv:hep-th/0404062].
}

\lref\iss{
  K.~Intriligator, N.~Seiberg and D.~Shih,
  ``Dynamical SUSY breaking in meta-stable vacua,''
  JHEP {\bf 0604}, 021 (2006)
  [arXiv:hep-th/0602239].
}

\lref\BenaRG{
  I.~Bena, E.~Gorbatov, S.~Hellerman, N.~Seiberg and D.~Shih,
  ``A note on (meta)stable brane configurations in MQCD,''
  arXiv:hep-th/0608157.
}

\lref\OoguriBG{
  H.~Ooguri and Y.~Ookouchi,
  ``Meta-stable supersymmetry breaking vacua on intersecting branes,''
  Phys.\ Lett.\ B {\bf 641}, 323 (2006)
  [arXiv:hep-th/0607183].
}

\lref\FrancoHT{
  S.~Franco, I.~Garcia-Etxebarria and A.~M.~Uranga,
  ``Non-supersymmetric meta-stable vacua from brane configurations,''
  JHEP {\bf 0701}, 085 (2007)
  [arXiv:hep-th/0607218].
}

\lref\ArgurioNY{
  R.~Argurio, M.~Bertolini, S.~Franco and S.~Kachru,
  ``Gauge / gravity duality and meta-stable dynamical supersymmetry breaking,''
  JHEP {\bf 0701}, 083 (2007)
  [arXiv:hep-th/0610212].
}

\lref\AganagicEX{
  M.~Aganagic, C.~Beem, J.~Seo and C.~Vafa,
  ``Geometrically induced metastability and holography,''
  arXiv:hep-th/0610249.
}
\lref\AhnGN{
  C.~Ahn,
  ``Brane configurations for nonsupersymmetric meta-stable vacua in SQCD with
  adjoint matter,''
  Class.\ Quant.\ Grav.\  {\bf 24}, 1359 (2007)
  [arXiv:hep-th/0608160].
}

\lref\AhnTG{
  C.~Ahn,
  ``M-theory lift of meta-stable brane configuration in symplectic and
  orthogonal gauge groups,''
  arXiv:hep-th/0610025.
}

\lref\HeckmanWK{
  J.~J.~Heckman, J.~Seo and C.~Vafa,
  ``Phase structure of a brane/anti-brane system at large N,''
  arXiv:hep-th/0702077.
}

\lref\TatarDM{
  R.~Tatar and B.~Wetenhall,
  ``Metastable vacua, geometrical engineering and MQCD transitions,''
  JHEP {\bf 0702}, 020 (2007)
  [arXiv:hep-th/0611303].
}

\lref\GiveonFK{
  A.~Giveon and D.~Kutasov,
  ``Gauge Symmetry and Supersymmetry Breaking From Intersecting Branes,''
  arXiv:hep-th/0703135.
}

\lref\KachruAW{
  S.~Kachru, R.~Kallosh, A.~Linde and S.~P.~Trivedi,
  ``De Sitter vacua in string theory,''
  Phys.\ Rev.\  D {\bf 68}, 046005 (2003)
  [arXiv:hep-th/0301240].
}

\lref\AcharyaRC{
  B.~S.~Acharya, K.~Bobkov, G.~L.~Kane, P.~Kumar and J.~Shao,
  ``Explaining the electroweak scale and stabilizing moduli in M theory,''
  arXiv:hep-th/0701034.
}

\lref\NakayamaGE{
  Y.~Nakayama, K.~L.~Panigrahi, S.~J.~Rey and H.~Takayanagi,
  ``Rolling down the throat in NS5-brane background: The case of  electrified
  D-brane,''
  JHEP {\bf 0501}, 052 (2005)
  [arXiv:hep-th/0412038].
}

\lref\SeibergEB{
  N.~Seiberg,
  ``Notes on quantum Liouville theory and quantum gravity,''
  Prog.\ Theor.\ Phys.\ Suppl.\  {\bf 102}, 319 (1990).
}

\lref\GaberdielZQ{
  M.~R.~Gaberdiel and A.~Recknagel,
  ``Conformal boundary states for free bosons and fermions,''
  JHEP {\bf 0111}, 016 (2001)
  [arXiv:hep-th/0108238].
}

\lref\AshokSF{
  S.~K.~Ashok, S.~Murthy and J.~Troost,
  ``D-branes in unoriented non-critical strings and duality in SO(N) and Sp(N)
  gauge theories,''
  arXiv:hep-th/0703148.
}

\lref\IsraelSI{
  D.~Israel and V.~Niarchos,
  ``Orientifolds in N = 2 Liouville theory and its mirror,''
  arXiv:hep-th/0703151.
}

%
\Title{\vbox{\baselineskip12pt\hbox{IC/2007/016}
}}%
{\vbox{\centerline{On supersymmetry breaking in string theory}
\vskip0.1cm
\centerline{from gauge theory in a throat}
}}

{\vskip -20pt\baselineskip 14pt   
\centerline{
Sameer Murthy$^{1}$ \footnote{}{$^1$\tt smurthy@ictp.it}  
}  
  
\bigskip  

\centerline{\sl Abdus Salam International Center for Theoretical Physics}  
\centerline{\sl Strada Costiera 11, Trieste, 34014, Italy.}  
\smallskip  
\bigskip\bigskip

\centerline{\bf Abstract}
We embed the supersymmetry breaking mechanism in $\CN=1$ SQCD of hep-th/0602239 in a smooth superstring theory using D-branes in the background $\IR^{4} \times SL(2)_{k=1}/U(1)$ which smoothly captures  the throat region of an intersecting $NS5$-brane configuration. 
A controllable deformation of the supersymmetric branes gives rise to the mass deformation of the magnetic SQCD theory on the branes. The consequent instability on the open string worldsheet can be followed onto a stable non-supersymmetric  configuration of D-branes which realize the metastable vacuum configuration in the field theory.
The new brane configuration is shown to backreact onto the background such as to produce different boundary conditions for the string fields in the radial direction compared to the supersymmetric configuration. In the string theory, this is interpreted to mean that the supersymmetry breaking is explicit rather than spontaneous.  
\noindent   
}

\Date{}  


\newsec{Introduction and summary}

Supersymmetry breaking at low energies is one of the most important issues to understand both for  supersymmetric models of particle physics and for superstring theory as a theory of quantum gravity if these  
are indeed the theories describing the real world. 
Recently, there has been some progress \iss\ in understanding a very generic phenomenon in minimal supersymmetric four-dimensional gauge theories, namely the existence of metastable vacua. The authors of \iss\ demonstrated that, under some well motivated assumptions, one can compute with control the lifetime of these vacua in $\CN=1$ SQCD, and they can be made parametrically large. 

There have been many papers following this work  \refs{\OoguriBG\FrancoHT\ArgurioNY\AganagicEX\HeckmanWK\AhnTG\AhnGN\TatarDM-\GiveonFK }
to embed and analyze this and related  field theory configurations in string theory, which when done successfully leads to the existence of such metstable vacua in string theory.\foot{There has been of course a lot of independent work {\it e.g.}\refs{\KachruAW ,  \AcharyaRC }  on finding and analyzing  non-supersymmetric metastable vacua in string theory; the advantage of an embedding described above would be that one has a simple but detailed understanding of the mechanism of supersymmetry breaking in the four-dimensional theory.} 
Since one is dealing with an $\CN=1$ supersymmetric theory in four dimensions, the examples talked about typically deal with  D-branes in string backgrounds which have singularities  
-- the simplest being of the conifold type \AganagicEX\ or a T-dual configuration of intersecting branes \OoguriBG . 

These singularities needs to be smoothed out; the examples studied so far were analyzed by working with either geometrically resolved/deformed configurations as in \AganagicEX , or in a region of parameter space where the fluctuations of the branes sitting at the singularities is still controllable and one can {\it e.g} use the DBI action for the branes \GiveonFK , or in a manifestly smooth M-theory lift \BenaRG . 

In this note, we shall demonstrate the existence of these stable non-supersymmetric vacua in a region of parameter space where one zooms in near the singularity keeping only the modes which are supported at the singularity; in this limit the background begins to develop a throat region\foot{For other work in finding stable vacua without supersymmetry in purely closed string backgrounds of a similar type, see \HarmarkSF .}. This is a deeply stringy regime and is relevant to the study of the decoupled field theory on the branes. 
One can perform a double-scaling limit where the throat is capped off and there is a finite string coupling at the tip \GiveonPX  ; in this situation,  one can use the methods of  conventional string perturbation theory.   The nature of our setup {\it i.e.} exact SCFT $+$ boundary states with a spectrum of gauge fields on them, ensures that  the assumption of \iss\ -- that the kinetic terms of the fields in the gauge theory are $O(1)$ -- is borne out in a manifest way. 

\subsec{The cigar setup as a limit of the ten-dimensional setup}

To be more precise, we shall study the background $\IR^{4} \times SL(2)_{k=1}/U(1)$ (cigar) which smoothly captures the throat region of the intersecting $NS5$-brane configuration with four flat common directions [See Figs 3-4], or equivalently that of the conifold in the double scaling limit mentioned above. This background has $\CN=2$ supersymmetry in four dimensions and has one modulus $\mu$ which encodes the string coupling at the tip. The string coupling decays exponentially away from the tip towards zero. The D-branes in the system  break a further half of the supersymmetry to $\CN=1$. 

The ten-dimensional type IIA intersecting brane configuration fig [3] from which our system descends has a parameter $L$ [see e.g. \BenaRG ] which is the length of the $D4$ branes stretched between the $NS5$-branes as well as $g_{s}$ which is the asymptotic string coupling constant. 
The branes could be treated as straight and the fluctuations could be treated in a Born-Infeld approximation when  $L/l_{s} >>1$, and $g_{s} \to 0$. To take into account effects of string interactions, the authors of \BenaRG\ instead studied the limit $g_{s} l_{s} \to \infty$ where one can use low energy techniques from M theory. To decouple the field theory on the branes, one needs to take the limit $g_{s} \to 0, L \to 0, l_{s} \to 0$ with the effective gauge coupling ${1 \over g^{2}_{YM}} \equiv {L \over g_{s} l_{s}}$ held fixed. 

The limit we take where we have an exact SCFT on the cigar is  $\mu \equiv {1 \over g^{tip}_{s}} \equiv {L \over g_{s} l_{s}}$ is fixed and $l_{s}$ is finite. To obtain the decoupled field theory on the branes, we then need to go to very low energies on the branes. 
In this limit, one can study the classical vacuuum states as configurations of D-branes and the fluctuations around them as open strings on the branes. 
Indeed, by doing so, one can recover the picture of Seiberg duality in this setup \murtro .

\subsec{D-branes which realize $SQCD$ and its deformations}

The D-branes in this background which break half the supersymmetry and realize $\CN=1$ SQCD on their worldvolume were studied in \refs{\FotopoulosCN , \Ashok , \murtro }.
The supersymmetric branes wrap the cigar and are labeled by the quantum numbers $(J,M)$ of the extended $\CN=2$ symmetry algebra on the worldsheet  \refs{\EguchiIK\IsraelJT\hoso\AhnTT\AhnQB\IsraelFN\RibaultSS-\FotopoulosUT } -- the characters and modular transformations of this algebra are summarized in Appendix A. 

The branes based on the identity representation $\ket{D3}$ have, at low energies, exactly the spectrum of the $\CN=1$ gauge theory on their worldvolume. These branes are localized at the tip of the cigar. The branes which realize flavor in the gauge theory extend to infinity, they are based on the continuous representation  $\ket{D5, J=M=\half}$ (electric flavor branes) and $\ket{D5,J=M=0}$ (magnetic flavor branes). 
%
%
These flavor branes realize massless quarks and mesons in the supersymmetric electric and magnetic configurations Figs [1,2]. Seiberg duality in this setup is realized as a partial monodromy $\mu \to -\mu$ which changes the basis of fundamental flavor branes from $\ket{D5, J=M=\half}$ to $\ket{D5,J=M=0}$ \murtro .  

The supersymmetric deformation in the field theory which corresponds to giving a mass to the quarks in the electric theory is understood as moving from the point  $(J=M=\half )$ onto the continuous branch $(J=\half + iP, M=\half )$ [Fig 1].  In the magnetic theory, the mass deformation breaks supersymmetry in perturbation theory and we present here a new corresponding deformation of the $(J=M=0)$ boundary state. Intuitively, this is done by trying to bend the color and flavor branes with respect to each other near the tip\foot{This is in the T-dual theory with momentum condensate where the branes are one-dimensional. On the cigar, this would correspond to turning on a small B-field on the extended brane near the tip.} Fig [7]. 
As in flat space, this breaks supersymmetry and the spectrum is tachyonic. 

The tachyonic quarks can then condense to get a non-supersymmetric stable minima. We can understand this on the open string worldsheet by another deformation of the unstable extended brane in which it swallows the localized brane and moves back a little from the tip. 
For both the deformations, we find that the spectrum of particles and symmetry breaking\foot{Note however that although the boundary state technology which we use is very useful to understand in detail the shape and couplings of a single brane, it does not teach us about the non-abelian nature of the gauge theory. Since we have a well-defined string perturbation theory, we use the standard approximation of thinking of a state of $N$ branes as a superposition of $N$ single boundary states. The symmetry breaking and the Goldstone particles related to the non-ablelian gauge groups cannot therefore be studied in detail in this formalism.}
is exactly what we expect from \iss . 

\subsec{Spontaneous v/s explicit SUSY breaking}

Finally, the explicit wavefunctions that we find for the non-supersymmetric state help us answer a theoretical question about supersymmetry breaking in this example. Semiclassically, the theory
of the closed strings and branes is defined by the configuration of closed and open string fields turned on at infinity\foot{ This statement is simply the assumption of a conventional notion of perturbative quantum gravity and is as such independent of holography. A boundary holographic theory in our linear dilaton background, being related to Little string theory \AharonyUB\  is not known to admit a conventional Lagrangian descrption unlike theories in AdS space. We still think of the boundary value of the non-normalizable bulk fields as coupling to the sources of the boundary theory. }. 
The state of the theory on the other hand is given by the shape of the normalizable fields in the bulk. We have two vacua, one supersymmetric\foot{It is true that we have a configuration for the classical supersymmetric vacuum, but we assume that the quantum fluctuations which produce the true supersymmetric vacuum are normalizable.} and one not, and one can ask if the two theories in which the vacua are defined are the same or not. 

In our case, there are two closed string fields which are relevant at infinity -- the tachyon winding mode and the RR axion which are related by the supersymmetry transformations of the closed string background at infinity \sameer .
The branes which preserve $\CN=1$ supersymmetry change the value of these two fields equally in the asymptotic region. On the other hand,  the backreaction of the stable non-supersymmetric configuration changes the two fields by unequal amounts. We interpret  this as saying that the supersymmetry breaking is actually occurring in the UV of the theory, and is explicit instead of spontaneous. A similar phenomenon  happenes in the M-theory lift of the same configuration \BenaRG\  due to brane-bending.

This is a little surprising since the gauge theory for which the susy breaking was spontaneous seemed to be localized in the bulk. From the point of view of the open string theory on the branes, this means that the UV completion of the  theory on the magnetic branes is qualitatively different from the asymptotically free  electric $\CN=1$ SQCD.  The speculation in \murtro\ that the full theory may include an effective  quartic interaction in the quarks may be relevant in this regard. 

\vskip 0.2 cm

\ndt {\it Note:} In the recent publication \GiveonFK , it was suggested in the context of related theories that dynamical transitions allowing changes in boundary conditions at infinity should be allowed. The analysis in this note may be useful towards understanding such time-dependent processes in a stringy regime, but we shall not discuss such transitions. Our above conclusion about supersymmetry breaking in string theory is drawn from comparing the asymptotic field configurations of  two {\it static}, {\it perturbatively stable} configurations -- the BPS brane configuration which at low energies realizes the supersymmetric vacuum in the $\CN=1$ SQCD theory near the tip {\it v/s} the configuration which realizes the metastable vacuum of the same field theory.

\vskip 0.2cm

The plan of the paper is the following. 
In section 2, we review the closed superstring background and its supersymmetric branes which realize  $\CN=1$ SQCD on their worldvolume. We then review how Seiberg duality in this gauge theory is realized as a partial monodromy in the closed string K{\"a}hler moduli space. 
In section 3, we exhibit a certain open string deformation of the magnetic theory which breaks supersymmetry and makes some of the quark directions tachyonic. Then, we exhibit another open string deformation which makes the theory perturbatively stable and corresponds to the metastable vacuum of \iss . Finally in section 4, we compute the backreaction of the metastable brane configuration onto the open and closed string fields and show that it differs from the supersymmetric one.

\newsec{Review of supersymmetric vacua of $\CN=1$ SQCD embedded in non-critical superstrings }

\subsec{Closed string background}

We consider the supersymmetric closed string background 
\eqn\backgnd{
\IR^{3,1} \times SL(2,\IR)_{k=1}/U(1).  
}
The worldsheet theory has $\CN=2$ supersymmetry, and a dual description of the coset theory is  $\CN=2$ Liouville theory. The coset theory has central charge $c=3+{6 \over k}$ and so the above background has $c=15$. Gauging the worldsheet $\CN=1$ supergravity and adding the usual $(b,c, \beta, \gamma )$ ghosts makes this a consistent string theory\foot{For a review, see {\it e.g.} \sameer .}. 

Semiclassically, the target space looks like $\IR^{3,1} \times$ a semi-infinite cigar with non-trivial metric and dilaton. We can parameterize the cigar by a radial coordinate $\rho \in [0,\infty)$  and angular coordinate $\theta \in [0, 2 \pi R)$ with $R=\sqrt{2k}$ in string units\foot{We shall set $\alpha'=2$ everywhere.}. To get $\CN=2$ worldsheet supersymmetry, we need to add two fermions  $\psi^{\pm} = \psi_{\rho} \pm i \psi_{\theta}= e^{\pm i H}$. 
Asymptotically as $\rho \to \infty$, this becomes simply a flat cylinder with the dilaton  linearly growing towards the tip. 
In the full CFT, the strong coupling region is capped off by the cigar tip, or in the dual description by the the sine-Liouville wall $S_{int} = \int d^{z} \psi^{+} \T \psi^{+} e^{-{1 \over Q}((\rho + \T \rho) + i(\theta - \T \theta))} + c.c. $. It is clear in either of these descriptions that the momentum around the cylinder is preserved and the winding is broken. There is one modulus $\mu$ which is related to the value of the string coupling at the tip as $\mu^{-1} = g^{tip}_{s}$.

The chiral part of the closed string vertex operators are labeled by $(j,m,s)$ which asymptotically have the wavefunction $e^{-j \rho + i (m +s) \theta + i s H}$. 
$(j,m)$ can take values corresponding to the continuous, discrete and finite representations of the $\CN=2$ algebra\foot{These are related to the representations of the $SL_{2}/U(1)$ coset, see {\it e.g.} \IsraelJT . }  on the worldsheet. $s$ labels the amount of spectral flow  of the $\CN=2$ algebra on the worldsheet. 
The state $\ket{j,m,s}$ has conformal weight and $U(1)_{R}$ charge 
\eqn\wtrcharge{
h = {(m+s)^{2} -j(j-1) \over k} + {s^{2} \over 2}, \qquad Q = {2(m+s) \over k} +s
}
and is annihilated by $G^{\pm}_{r},r \ge \half \mp s $. $s=0$ is the NS sector primary state, and $s=\half$ is the R sector.

The irrational nature of the conformal field theory brings in subtleties compared to rational CFT's, some of which have been understood well. 
In particular, to have nice properties such as integral $U(1)_{R}$ charges under modular transformations, one needs to consider the characters of the extended $\CN=2$ algebra, which includes the spectral flow generators \EguchiIK . In Appendix A, we list the various characters and their modular transformations. 

The spacetime supersymmetry is two copies of $d=4$ Poincare supersymmetry (one each from the left and the right movers on the worldsheet). The momentum around the cigar is a $U(1)_{R}$ symmetry.

\subsec{The supersymmetric branes which realize $\CN=1$ SQCD\foot{This subsection is simply a review of \murtro .} }

We shall consider D-branes which fill the $\IR^{4}$ directions in order to discuss $d=4$ gauge theories realized on them. On the cigar, the supersymmetric branes preserve the classical $U(1)_{R}$ momentum symmetry, and a diagonal combination of the left and right moving supercharges. 
There are two types of branes possible -- the first type which has no modulus is localized near the tip and has a tension proportional to $\mu$. The only massless modes on its worldvolume are those of a $\CN=1$ gauge multiplet. The other type of brane extends along the radial direction; asymptotically it  can be understood as a Neumann brane in the linear dilaton background. This type of brane semiclassically is labeled by a parameter $\mu_{B}$ which can be thought of as a boundary cosmological constant which indicates where the brane dissolves towards the strong coupling region.

In the exact SCFT,  the boundary states are built using a Cardy-type ansatz.  This gives us boundary states corresponding to each of the representations discussed above for various values of the parameters\foot{The range of parameters could in general be different from the closed string theory because of the irrational nature of the SCFT. In the absence of the Verlinde formula and a Cardy solution, we should allow for all possible values of the parameters based on the formal algebraic  reps above which have consistent open + closed string physics.}. 
We first define the Ishibashi states:
\eqn\ishi{
\bbra{j,m} e^{- \pi T H^{cl}}e^{i \pi z(J+\T J)} \kket{j',m'} =  \left( \delta(j-j') + R(j) \delta(j+j'-1) \right)\delta_{\IZ_{2}} (m,m') Ch_{j, m}( i T, z).
}
where $R(j)$ is the reflection coefficient from the tip of the cigar; 
and look for solutions to the following equations:
\eqn\cardy{\eqalign{
\bra{B;Id} e^{- \pi T H^{cl}}e^{i \pi z(J+\T J)} \ket{B;Id} & =  Ch_{Id}^{NS}(i t, z'), \cr 
\bra{B;Id} e^{- \pi T H^{cl}}e^{i \pi z(J+\T J)} \ket{B;\xi} & =  Ch_{\xi}^{NS} (i t, z'), \cr
T \equiv {1 \over t}, \; z' = - i t z. &\cr
}}
We have used the notation $T$ for the closed string channel modulus, and the
notation $t$ for its open string annulus counterpart.

\ndt The solutions are as follows. The identity brane has the one-point functions ($\CT \equiv \half + i \IR$):
\eqn\idbrane{\eqalign{
\ket{B;Id} & = \int_{\CT} {dj} \sum_{m = 0,\half} \Psi_{Id} (j,m) \kket{j,m} \cr
& \Psi_{Id}(j,m) =  \nu^{j-\half} {\G(j + m) \G(j - m) \over \G(2j-1) \G(2 j) }. \cr
}} 
and the brane associated to the continuous representation is:
\eqn\contbrane{\eqalign{
\ket{B; cont, J,M} & = \int_{\CT} {dj \over 2\pi} \sum_{m = 0,\half} \Psi^{cont} (j,m) \kket{j,m} \cr
\Psi^{cont}(j,m) &
 = (2 \pi)^{1 \over 2}  \mu^{j-\half-m} {\bar \mu}^{j-\half + m}   \cos{(4 \pi (J-\half) (j-\half))}
{\G(1-2 j) \G(2- 2 j)    \over \G(1 -j + m) \G(1- j - m) }
e^{4 \pi i M m}. \cr
}}

To build a four dimensional field theory, we need to tensor the profiles in the extended four dimensions. 
We choose all the branes to be Neumann in the $\IR^{1,3}$ direction, and they have different profiles on the cigar.
The supersymmetric branes relevant to $\CN=1$ SQCD with $SU(N)$ gauge group are based on the identity representation and the continuous representation and were denoted $\ket{D3}$ (color) and $\ket{D5;J,M}$ (flavor). We choose a configuration of $N_{c}$ color branes and $N_{f}$ of the flavor branes. It is also possible to engineer theories with $SO/Sp$ gauge groups by using orientifolds in these backgrounds \refs{\AshokSF ,  \IsraelSI }.

The low energy theory on the $\ket{D3}$ is an $\CN=1$ pure $SU(N_{c})$ gauge theory with gauge fields $A^{ab}$. The $\ket{\bar{D5;J,M=\half}}$\foot{The presence of the bar indicates the sign of  charge of the branes under the RR axion which winds around the cigar; for now we only discuss branes that are mutually supersymmetric with each other and with the $\ket{D3}$.} introduces left and right handed quark multiplets $Q^{ai}, \T Q^{i}_{a}$  $(a=1,..,N_{c}, i = 1,..,N_{f})$ with mass $m^{2} = -J(J-1) - {1 \over 4} $ into the gauge theory. These fields are open strings stretched between the color and flavor branes, and so they are also localized near the tip. 
The $\ket{D5;J,M=0}$ introduces massive left and right handed quark multiplets $q^{ai}, \T q^{i}_{a}$ into the gauge theory with mass $m^{2} = -J(J-1)$.\foot{At $J=0$, its overlap with the $\ket{D3}$ also shows the presence of gauge fields with quantum numbers $A^{ia}_{\mu}$, these become massive and are discussed below.}  The allowed values of $(J,M)$ are $(J \in \CS = \half + i \IR^{+}, M=0,\half )$, and  $(0 \le J \le \half, M=0)$. 

\ifig{\cigelecmass}{Supersymmetric branes on the cigar $\ket{D3}$ and $\ket{\bar{D5,J=\half + iP, M=\half}}$ realize electric SQCD with massive quarks. The extended branes are double sheeted and dissolve before reaching the tip  at a distance  which determines the mass of the quarks.}{\epsfxsize3.0in\epsfbox{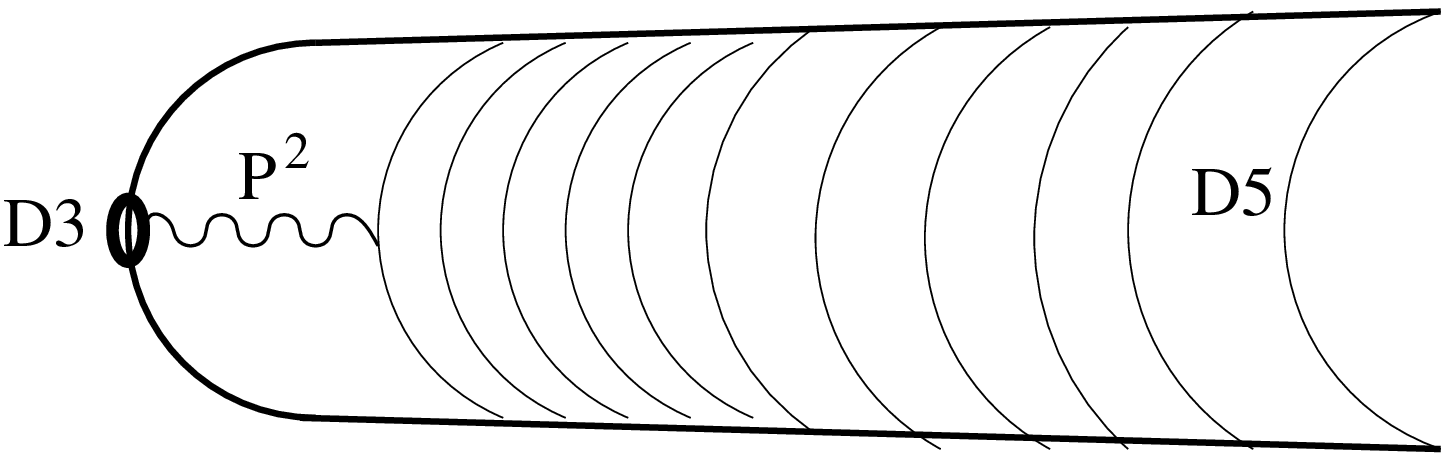}}

In their self-overlap, the branes with $M=\half$ do not introduce any massless four dimensional modes. For  $M=0$, there is no new massless four dimensional mode in the self-overlap for any value of $J \neq 0$. At $J=0$, one finds a massless meson $M^{ij}$ charged in the adjoint under the $SU(N_{f})$ rotating the flavor branes.

\ifig{\cigJbranch}{Supersymmetric $\ket{D5,J,M=0}$ branes on the cigar with $(0 \le J \le \half)$. These branes fill the cigar and have a unit of magnetic flux localized near the tip which have different profiles determined by $J$. At $J=0$, the flux concentrates at the tip and the configuration becomes $\ket{D5,J=M=\half} + \ket{D3}$.  These realize magnetic SQCD with no mass deformation.}{\epsfxsize3.0in\epsfbox{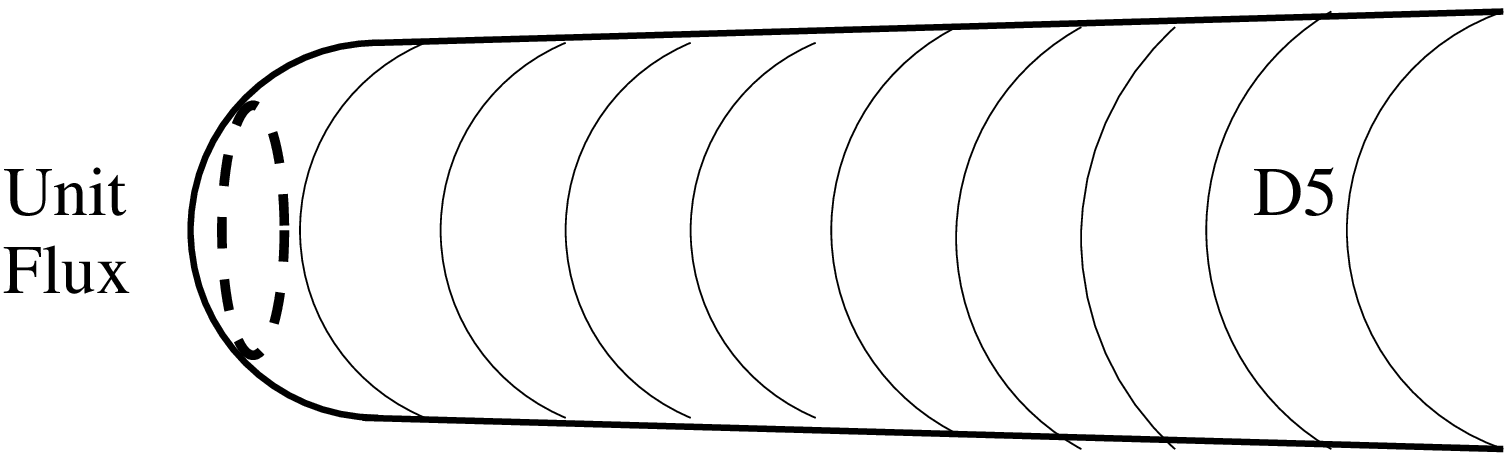}}

The configuration of $N_{c}$ $\ket{D3}$ branes and $N_{f}$ $\ket{\bar{D5,J=M=\half}}$ gives rise to the electric picture of $\CN=1$ SQCD with massless quarks. In this configuration, there is a brane addition relation based on \charaddn :
\eqn\braneaddn{
\ket{\bar{D5, J=M=\half }} + \ket{D3} = \ket{D5, J=M=0}.
}
The magnetic configuration is achieved by the partial monodromy $\mu \to - \mu$, which induces the following transformations\foot{The negative sign in front of the branes is necessary to keep the tension  $\mu$ positive.} on the branes: $\ket{D3} \to (-\ket{D3}), \, \ket{\bar{D5,J=M=\half}} \to \ket{D5, J=M=\half}$. 
After tachyon condensation on $N_{c}$  $(-\ket{D3})$ and $(-\ket{\bar{D3}})$, we are left with a configuration of $\T N_{c} = N_{f} - N_{c}$ $(-\ket{\bar{D3}})$ branes and $N_{f}$ $\ket{\bar{D5,J=M=0}}$ 
such that the total charge of the branes before and after the monodromy remains the same.
The addition relation above now is reexpressed as
$\ket{\bar{D5, J=M=0 }} + (-\ket{\bar{D3}})= \ket{D5, J=M=\half}$,
where all the branes have positive tension. 

This configuration has the fields of the magnetic dual of $\CN=1$ SQCD with massless magnetic quarks and mesons. There are also the gauge fields $A^{ia}$ mentioned above which are massless at tree level. However, the global rotation and R charges of the quarks and mesons do not allow a gauge interaction with these fields and they do not appear in the four dimensional low energy description. This is interpreted as the fact that the interactions will generate a mass for these fields.\foot{This is consistent with the fact that, unlike the color gauge fields, it is not clear whether there is a BRST trivial open string operator which provides the decoupling of the longitudinal modes. It is also consistent with the ten-dimensional Hanany-Witten type setups which we discuss below.}
There is a superpotential coupling of three open string fields allowed by the global rotation symmetries and classical R symmetry of the theory $W=q M \T q$. We can summarize by saying that after the monodromy, we are left with the magnetic SQCD. 

We can identify our non-critical brane setup with the near-horizon limit of the ten-dimensional type IIA setup \ElitzurHC\ involving orthogonal $NS5$ branes, $D4$-branes suspended between then, and $D6$-branes stretched to infinity [Figures 1 and 2 below]. We shall use the conventions of \BenaRG\ for the brane setups in ten dimensions.

\ifig{\elec}{Supersymmetric Electric configuration with massless quarks. The setup in ten dimensions follows the conventions of \BenaRG .}{\epsfxsize2.0in\epsfbox{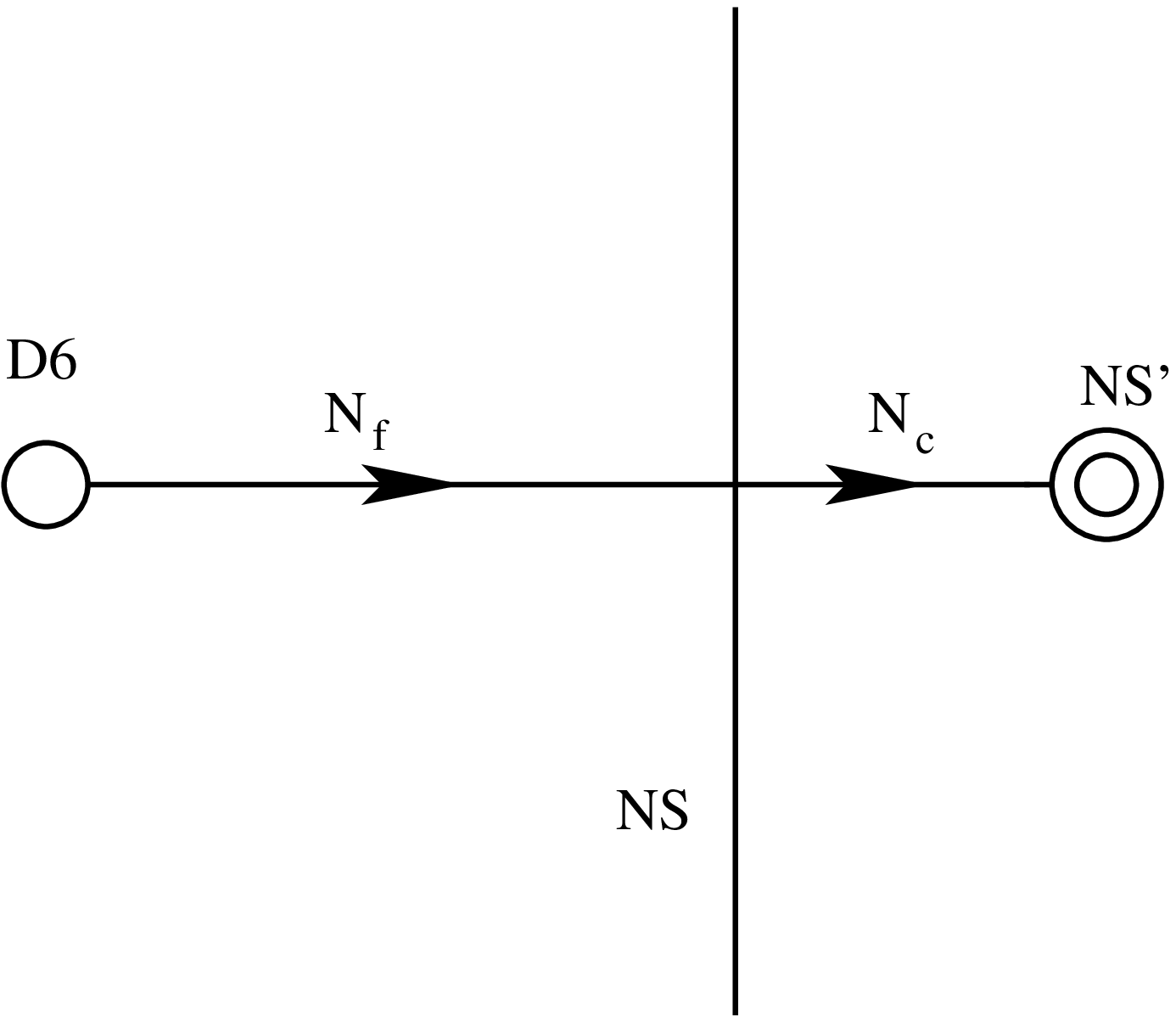}}

\ifig{\elec}{Supersymmetric Magnetic configuration with massless quarks.}{\epsfxsize2.0in\epsfbox{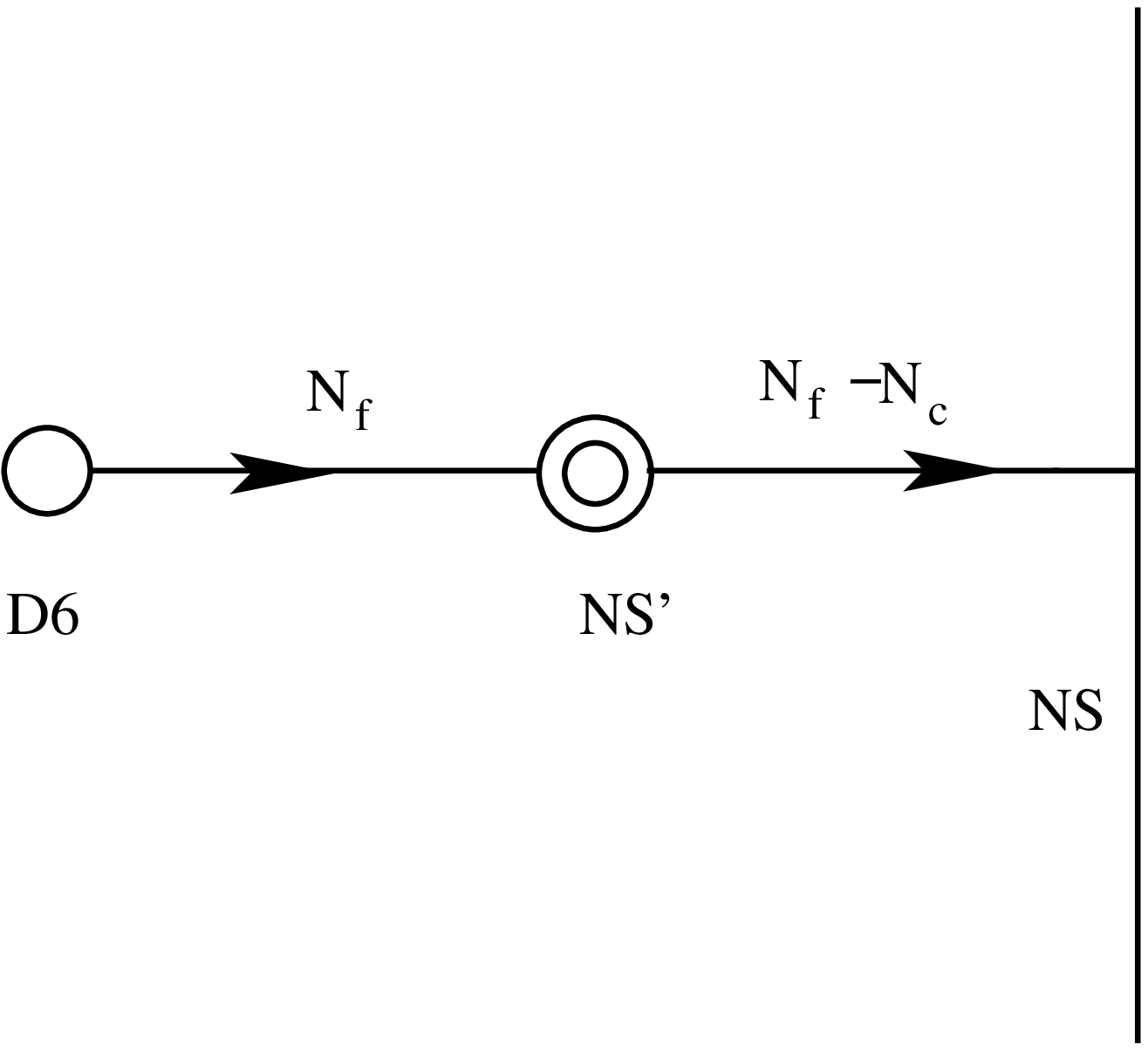}}

The $NS5$-branes in the figures, are replaced by our type IIB closed string background \backgnd . 
The $\ket{D3}$ branes are identified with the $D4$-branes in the figure.
The $D4$-branes suspended between the two $NS5$-branes has one of its directions spanning an interval of length $L$ -- this parameter enters the gauge coupling as $g_{YM}^{-2}= (L/l_{s}) g_{s}^{-1}$. The gauge theory is effectively four dimensional for small values of $L/l_{s}$. In our setup, we have only one parameter $\mu$ which controls the gauge coupling. 

The $\ket{D5,J=\half}$ branes are identified with the $D6+D4$ brane where the $D4$ is  stretched between the $D6$ and $NS5$. This $D4$ is rigid and thus there is no localized four dimensional fluctuation on the extended brane. This brane gives rise to flavors in the electric setup. 
The $\ket{D5,J=0}$ branes are identified with the $D6+D4$ brane where the $D4$ is  stretched between the $D6$ and $NS5'$. This $D4$ has a massless scalar which corresponds to its motion in the parallel directions of the two extended objects. This is the magnetic flavor brane, and has the four-dimensional meson mode localized near the intersection. 

The addition relation \braneaddn\ is simply the fact that the $D4$-brane between the two $NS5$-branes can add to the first type of $D6+D4$ to become the second type as is clear from the picture. 
In terms of these pictures, Seiberg duality was realized at a classical level by moving the two $NS5$-branes around each other and showing that the various $D$-branes rearrange themselves in such a way to reach the magnetic configuration [Fig. 2] starting from the electric one  [Fig. 1]. This could be done smoothly by turning on an $FI$ parameter which avoids any singular points in such a monodromy. 

This picture was good enough to understand F-term information in the gauge theory like the moduli space of vacua and the chiral rings. The non-critical setup in principle has more information because of the exact nature of the background. In the following, we shall exploit it in one such direction by turning on deformations which break supersymmetry.

\newsec{New Branes relevant to broken supersymmetry}

In the electric theory, the mass deformation which corresponds to adding a superpotential $W = {1 \over 4} {\rm Tr} \, \a \, Q \T Q$ is supersymmetric [fig 5]. This is understood as the change of the parameter $J= \half + iP$ in this continuous branch away from $P=0$ to $P= \a$. 

\ifig{\elecmass}{Supersymmetric electric configuration with massive quarks}{\epsfxsize2.0in\epsfbox{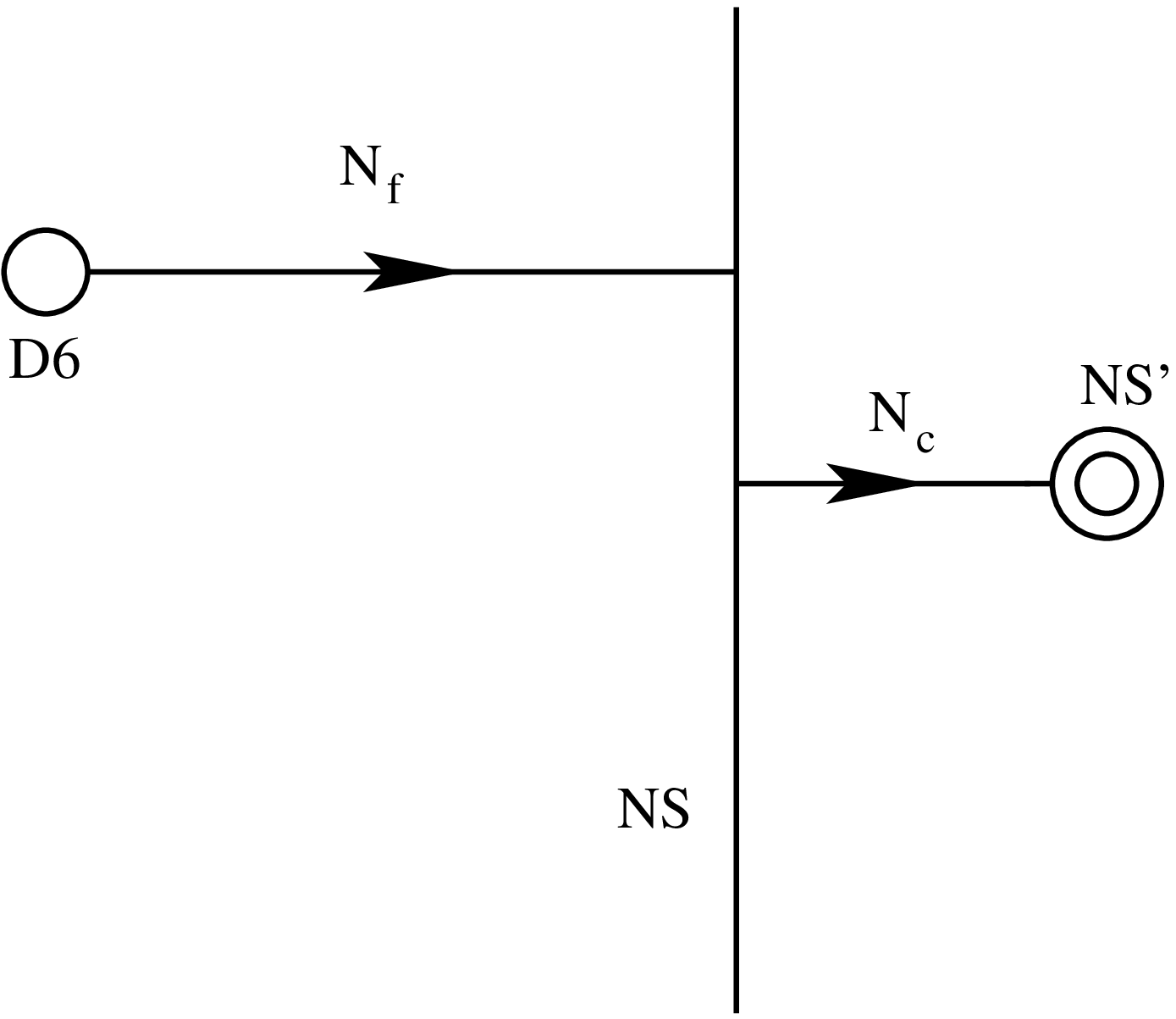}}

\subsec{Unstable Brane configurations which spontaneously break supersymmetry}

In the magnetic theory, this deformation corresponds to adding a term linear in the meson 
\eqn\superpotmag{
W = {\rm Tr} \, q M \T q + \a {\rm Tr}\, M . 
}
As was shown in \iss , for $N_{f} > N_{c}$, this deformation breaks supersymmetry spontaneously at tree level simply because there are more $F$-term equations than fields. 
Expanding the scalar potential for the above superpotential, we find that there is a quadratic term for the fields $q$ and $\t q$ with coefficient $m^{2} = \langle M^{2} \rangle- \a$. For small values of $\langle M^{2} \rangle$, this implies a complex tachyon. 
In the brane setup in the type IIA theory \refs{\OoguriBG , \BenaRG }, a deformation causing such an instability is implemented by a certain deformation of the magnetic flavor brane as shown in Figure 4. 

\ifig{\magmass}{Magnetic configuration with mass deformation. This configuration breaks supersymmetry and some of the quark directions become tachyonic.}{\epsfxsize2.0in\epsfbox{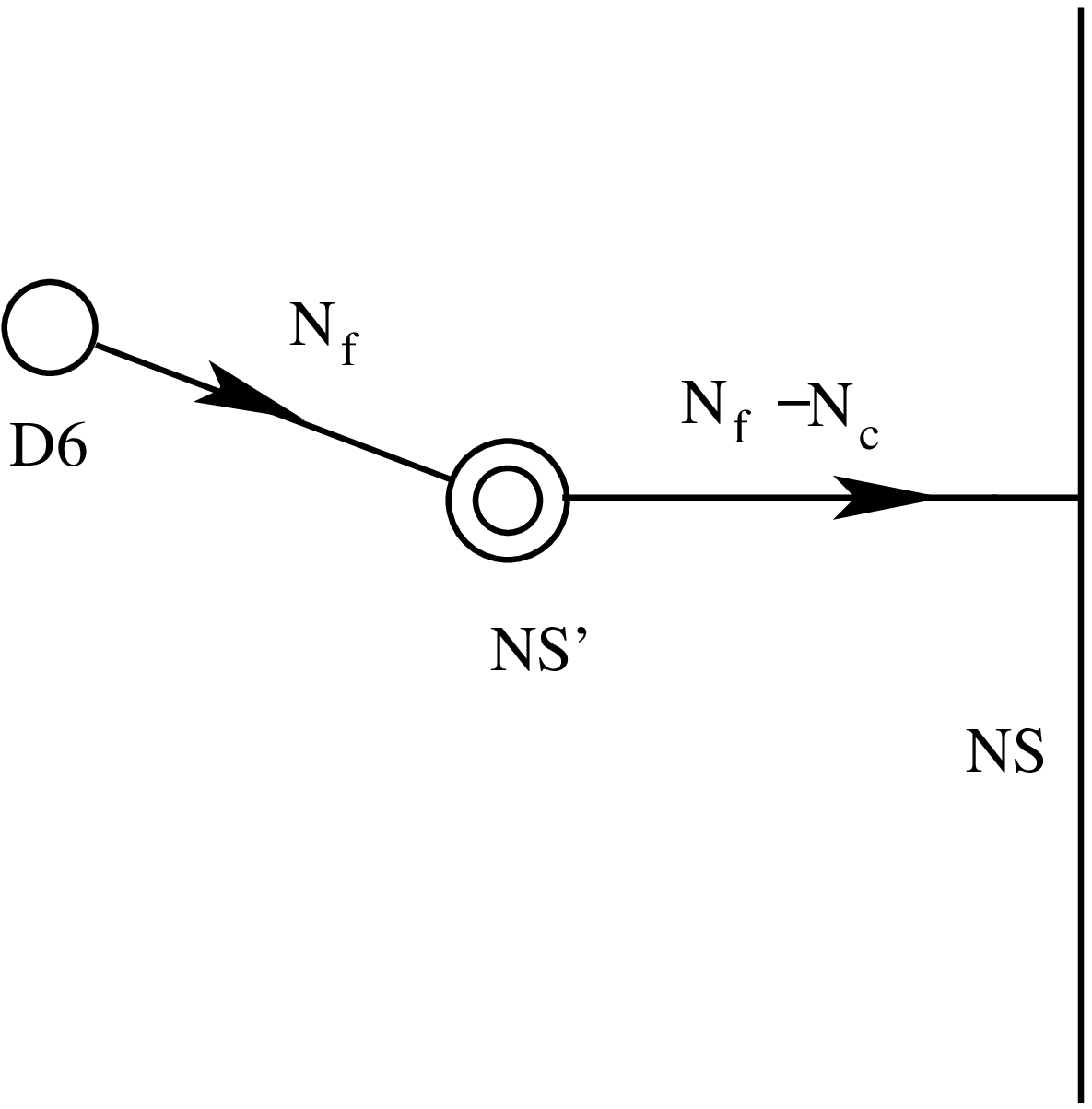}}

We look for deformations of the boundary states $\ket{D5; J=M=0}$ such that the new flavor brane has a self-overlap which is the same as that of the undeformed flavor brane and hence at tree level has a massless meson multiplet -- this implies that $J$ must remain zero. This brane however must break supersymmetry completely and in the overlap with the color branes, one of the quark degrees of freedom become tachyonic. Such a deformation must of course preserve $\CN=1$ supersymmetry on the worldsheet which is gauged.

In the sine-Liouville theory (where the branes under consideration are one-dimensional objects going in from infinity towards the tip and turning back), we also have an intuition that the deformed brane is slightly bent around the circle away from the localized color brane (which also has two branches, but decays quickly towards infinity). This deformation of the flavor branes must however decay towards infinity (since that is so in the type II setup). 

\ifig{\tachyons}{A heuristic picture for the supersymmetry breaking branes. In the sine-Liouville picture on the cylinder with momentum condensate, the branes we have been considering have two legs, one coming in towards the strong coupling region, and the other going out at a diametrically opposite point on the circle. The double lined $D3$ branes are localized in the strong coupling region -- in the figure, they seem to be cut off going towards the weak coupling region, but they have a smooth profile as given by \idbrane . The single-lined $D5$ branes go out all the way to infinity. To break supersymmetry locally, we can try to bend the $D5$ branes with respect to the $D3$ branes, but only in  the strong coupling region. Such a deformation is captured by \deformchar\ below. As to whether this deformation really dies away towards the weak coupling region is answered in the following section.}{\epsfxsize3.0in\epsfbox{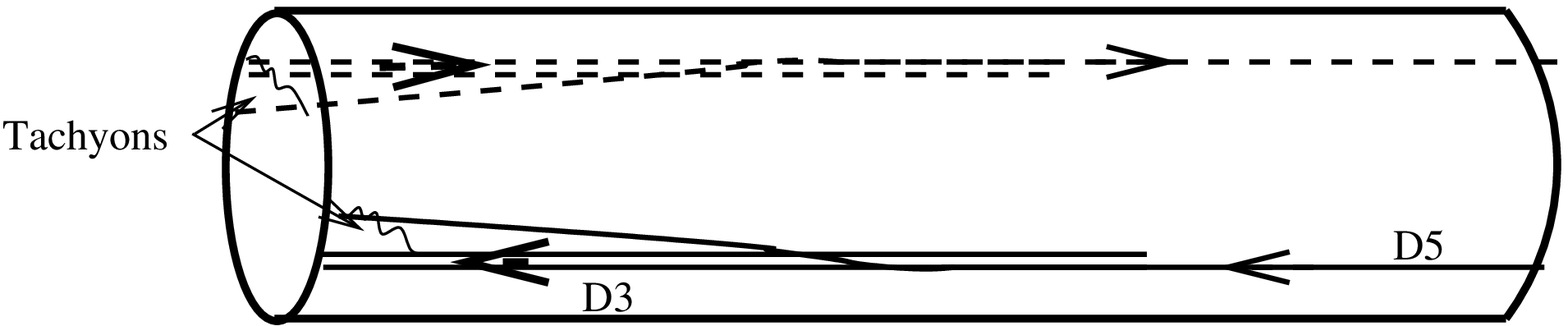}}

A deformation of the open string (NS) character $(J=M=0)$ obeying the above properties is given by:  
\eqn\deformchar{
Ch_{J,M}^{\a}(t,z) = q^{-J(J-1)-{1 \over 4}} \Theta_{2M,1}(t,z) {1 \over \cos(\pi \a)} {\sin(\pi \a)   \; \vartheta_{00} (\tau, z+ \a t)\over \vartheta_{11}(\tau, z+ \a t) }. 
}
Note that this deformed character can be interpreted as the following (see also \NakayamaGE ) -- the continuous representation \chars\ is built out of a free field module tensored with $q^{h-{1 \over 4k}}$ arising from the zero mode of the Liouville theory. At $h=0$, one can turn on a twist in the two directions exactly like branes at angles/branes with B field in flat space.\foot{
We shall not try here to analyze thoroughly the space of all possible $\CN=1$ preserving deformations of the $\CN=2$ Liouville boundary states. This space is quite large even for the free field system with $c=3$ \GaberdielZQ .} Such a twist is consistent with $\CN=1$ supersymmetry on the worldsheet. 
Then one must sum over the lattice of zero modes as before\foot{Note again that this is not a construction which respects  $\CN=2$ supersymmetry -- the $\CN=2$ spectral flow of the twisted free field character gives a lattice sum which is different from the one above. See \GaberdielNV\  for such a construction for $c=3$.}. This lattice sum is not touched by the twist and so also respects $\CN=1$ supersymmetry on the worldsheet as before.

We can now define deformed Ishibashi states which which obey the condition:
\eqn\newishi{\eqalign{
&\bbra{j,m,\a} e^{- \pi T H^{cl}}e^{i \pi z(J+\T J)} \kket{j',m',\a'}   = 2 \pi \left( \delta(j-j') + R(j) \delta(j+j'-1) \right)\delta_{\IZ_{2}} (m,m') \times  \cr
&  \qquad \qquad \times q^{-j(j-1)-{1\over 4}} \Theta_{2m,1}(T,z) {1 \over \cos(\pi \a) \cos(\pi \a')} {\sin{\pi (\a - \a')}   \vartheta_{00} (T, z+ \a - \a')\over \vartheta_{11}(T, z+ \a-\a'). }  \cr
}}
Note the  $\a$ dependent normalization -- this is the free field normalization for branes at angles/with $B$  flux [see e.g. \refs{\polchinski , \GaberdielNV}]. This normalization affects the one-point functions, and the one above is consistent with charge conservation of the snapping process which we shall talk about soon. 

\ndt Using the same one-point function as before, we define the Cardy states:
\eqn\newcontbrane{\eqalign{
\ket{B; J,M,\a} & = \int_{\CT} {dj} \sum_{m = 0,\half} \Psi^{cont} (j,m) \kket{j,m,\a} \cr
\Psi^{cont}(j,m) &
 =  \nu^{j-\half}   \cos{(4 \pi (J-\half) (j-\half))}
{\G(1-2 j) \G(2- 2 j)    \over \G(1 -j + m) \G(1- j - m) }
e^{4 \pi i M m}. \cr
}}
with the open string spectrum:
\eqn\newcardy{
 \bra{B;Id,0} e^{- \pi T H^{cl}}e^{i \pi z(J+\T J)} \ket{B;J,M,\a}   = Ch_{J,M}^{\a}(t,zt).
 }

\ndt Let's now deform the magnetic brane as above with parameter $\a$. So far, we discussed the NS sector amplitudes.  After adding in the other sectors, we find that 
the full spectrum of strings stretched between the color brane $\ket{\bar{D3}}$ and $\ket{\bar{D5;J=M=0,\a}}$ is:
\eqn\quarkspec{\eqalign{
& \bra{\bar{D3}} e^{-TH_{cl}} \ket{\bar{D5; J=M=0,\a}}  \cr
& =  V_{\a} {q^{-{1 \over 4}}  \over \vartheta_{11}(t, \a t) \eta^{3}(\tau) } 
\half  \left[  \Theta_{01}(t,0)
 \left( \vartheta_{0  0} (t,0) \vartheta_{0  0} (t,\a t)
 -  \vartheta_{0 1} (t,0) \vartheta_{0 1} (t,\a t)  \right) \right. \cr
& \qquad \qquad \qquad \left.  - \Theta_{1 1} (t,0)
 \vartheta_{1 0} (t,0)  \vartheta_{1 0} (t, \a t)  \right]  \cr
& =  V_{\a} \left[ \left( q^{-\a/2}  +  q^{\a/2}  + ...\right)  + 
\left(2 q^{\a/2} + ...\right) + \right. \cr
& \qquad \qquad \qquad - \left\{  \left. \left( 2 + ... \right)   + \left(2q^{\a} + ...\right)  \right\}\right]  \cr
}}
where the first line is the NS sector and the second is the R sector (in the open string channel). The overall normalization is the same one chosen for branes with flux in flat space, such that the expression is finite with pure volume factors as  $\a \to 0$ \polchinski . We have grouped the factors in parentheses such that the first two excitations come from the oscillators in the cigar directions, whereas the second two arise from the flat space oscillators. The former correspond to the quark excitations (there are two more coming from the complex conjugate diagram) and the latter to the gauge field, which as discussed earlier, picks up a mass due to interactions and decouples. The ellipses correspond to oscillators with higher masses. 

We can now read off the open string spectrum at low energies for $0<\a <<1$. While earlier, at $\a =0$, we had two massless quark superfields $q$ and $\T q$, the masses of all the modes shift in the presence of the deformation. Among the bosonic squark fields, we now have  two tachyonic modes $m^{2}= -\half \a$, two massive  modes with $m^{2} = +\half \alpha$; and all the fermionic quarks remain massless.  The localized self-overlap of this brane $\ket{D5;J=M=0,\a}$ does not depend on $\a$ and therefore contains as before a massless meson multiplet. The pattern of masses above is exactly that\foot{At $g_{s}=0$, we can understand the deformation on $N_{f}$ boundary states as performing the above deformation on each of the extended states independently.}
 which comes from the Lagrangian \superpotmag\ of the magnetic theory deformed by the mass parameter expanded around a configuration where the meson has not acquired a vev.

\subsec{New branes with perturbatively stable spectrum}

The magnetic SQCD configuration deformed by the small parameter $\a$  above has a tachyonic quark direction with mass $m^{2} = - \half \a$. This mode can condense and it has been argued \iss\  that there is a non-supersymmetric metastable vacuum at the end of the condensation. As shown in \iss , the vacuum energy in this state is proportional to $|\a|^{2}$ and the quark fields have an expectation value $q \T q$ proportional to $\a$. For small values of $\a$, this vacuum is close compared to the classical magnetic vacuum at $\a = 0$. 

In the IIA brane picture, this corresponds to  $N_{f}-N_{c}$ D4 branes stretched between the NS and NS' joining with an equal number of D4 branes stretched between the D6 and NS' and snapping back into a rigid brane stretched between the D6 and NS \BenaRG\ (fig. 8). This rigid brane has the same quantum numbers as  a flavor brane in the  {\it electric} configuration which gives masses to the electric quarks.

\ifig{\stable}{Stable non-supersymmetric  configuration. This configuration is reached when the deformed branes of Figure 4 snap back and relax into the minimum energy configuration. There are no perturbative instabilities and this configuration is long-lived.}{\epsfxsize2.0in\epsfbox{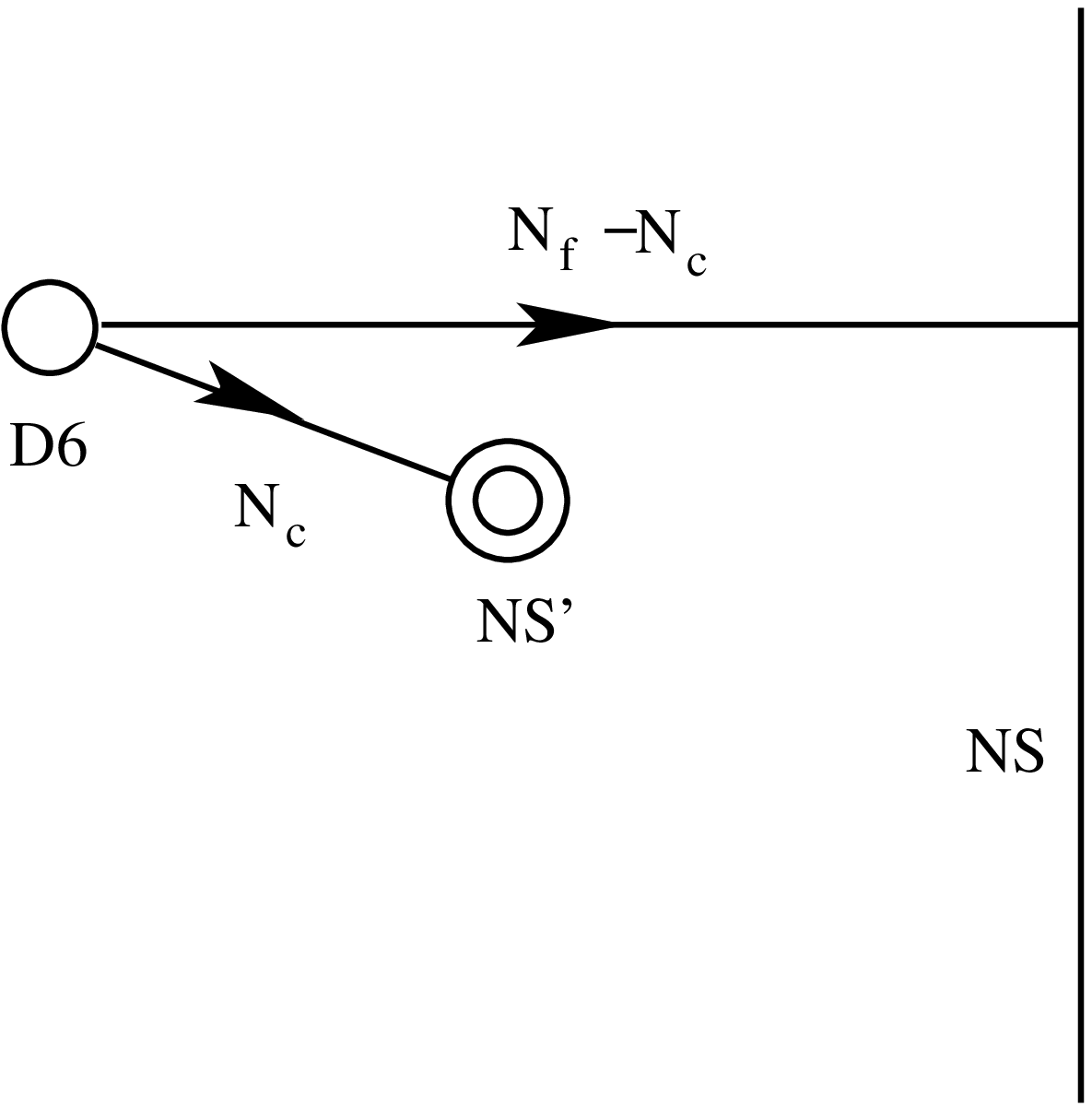}}

In the near-horizon limit captured by the non-critical theory, this snapping process in the magnetic configuration should be described by  the $\ket{\bar{D5; J=M=0,\a}}$ joining with the $(-\ket{\bar{D3}})$ to form a $\ket{D5}$. The new $D5$ brane is one that we have already discussed -- it is in the continuous representation with $P =  \a$  [Fig 5].

At $\a=0$, the charges of the three branes above were (in units of the $\ket{D3}$ charge), $-\half, +1, + \half$. On deformation, the charges of the second two branes remains the same, we will see soon that the charge of the deformed brane also stays the same, and hence the above snapping is allowed by charge conservation.

The overlap between two extended branes is given by:
\eqn\contlap{\eqalign{
& e^{\pi {3 z^{2} \over T}} 
\bra{B;J_{1},M_{1},\a } e^{- \pi T H^{cl}}e^{i \pi z(J+\T J)} \ket{B;J_{2},M_{2},\a' } \cr
 & = \int_{-\infty}^{\infty}  {dp} \left[ \rho_{1}(p|J_{1},J_{2}) Ch^{\a-\a'}(p,M_{2}-M_{1};it,z') \right.
 + \left. \rho_{2}(p|J_{1},J_{2})  Ch^{\a-\a'}(p,M_{2}-M_{1} +\half ;it,z') \right] \cr
}}
where the spectral densities $\rho_{i}$ are given by:
\eqn\specden{\eqalign{
& \rho_{1}(p | J_{1},J_{2})  =  \int_{0}^{\infty} {dp'} {\cos{(4 \pi p p')} \over \sinh^{2}{(2 \pi  p')} } \sum_{\epsilon_{i}=\pm 1} \cosh{\left(4\pi \left(\half + i \epsilon_{1} P_{1} + i \epsilon_{2} P_{2}\right) p' \right)}
\cr 
& \rho_{2}(p|J_{1},J_{2})  = 2 \int_{0}^{\infty} {dp'} {\cos{(4 \pi p p')} \over
 \sinh^{2}{(2 \pi  p')} } \sum_{\epsilon=\pm 1} \cosh{\left(4\pi \left( i
 P_{1} + i \epsilon P_{2}\right) p' \right)} .
\cr 
}}
For $J_{i}-\half \equiv P_{i} \in \IR^{+}$, these formulas are well-defined. For imaginary values of
 $P_{i}$, corresponding to $0 \le J \le \half$, one has
 to be more careful. The $p'$ integral in \specden\  may generate additional
 divergences at $p' = \infty$, which can
 be eliminated by shifting the contour of $p$ integration in \contlap\  
 before exchanging the order of the integrals [see \murtro ]. 
Thereafter, one can freely exchange the integral and shift back the contour,
and find if there any additional contributions to the brane spectrum. 

In our setup, we have at the end of the process $N_{c}$ $ \ket{J=M=0,\a}$ and $(N_{f} - N_{c})$ $\ket{J=\half + i  \a , M=\half,\a=0}$. The self overlaps of the brane $ \ket{J=M=0,\a}$ is independent of $\a$ and we know already that it does not have open string tachyons in its spectrum. This is simply the fact that \contlap\ gives us a massless meson $+$ massive modes on the supersymmetric magnetic flavor branes. Note that the meson remains massless at tree level in the $\a$-deformed theory as well. This is as in \iss  -- these pseudomoduli get a potential due to the interaction with the $(N_{f}-N_{c})$ flavor branes with $J$ in the continuous branch. We know that the brane $\ket{J=\half + i  \a, M=\half,\a=0}$ is stable as well. The only potential source of perturbative tachyons is the overlap between the two types of branes computed from \contlap  . 

In this overlap, we find that there are no additional localized contributions from outside the continuum.\foot{There is a contribution to $\rho_{1}$ which appears at the edge of the continuum as for the $J={1 \over 4}$ brane self overlap in \murtro . It appears not as a pole, but because of a logarithmic branch cut and is not counted as a genuinely localized mode. In any case, this mode is not tachyonic.}
After tensoring in the $\IR^{4}$, we find that the continuum of modes in the open string spectrum  \specden\  begins with  $m^{2}=0$  (NS,R) in $\rho_{1}$ and $m^{2} = {1 \over 4} - \half \a$ (NS) and $m^{2} = {1 \over 4} $ (R) in $\rho_{2}$. The conclusion is that for $\a <<1$, the configuration of these two types of branes is perturbatively stable. On the cigar, we can understand this by thinking of the new $N_{f}-N_{c}$ branes as having moved away from the tip a little distance proportional to $\a$ [Fig 1] causing the open strings stretching from the tip and ending on them to become non-tachyonic. 

\newsec{The Backreaction of the branes}

A relevant question is whether the deformation of the brane is truly localized or not. In terms of the stick-figure type IIA setup [Figures 1-4], there is a possibility for modes to hide in the singularity. The details of how string theory resolves this singularity could potentially affect the true answer. The advantage of using the exact CFT is that we can answer this type of dynamical question precisely in terms of conventional perturbative string theory.\foot{As mentioned in the introduction, another approach to resolve the singularity would be to lift the configuration to M theory as in \BenaRG .}
 In terms of the sine-Liouville picture presented earlier, the question is whether the bending of the extended brane (in the circle direction, with respect to the color branes) near the strong coupling region decays fast enough or not towards the weak-coupling region. 

In our setup, the  non-normalizable operators which define the closed string background are the tachyon winding mode with value $\mu_{ren}$ and the value of the RR axion (which is zero without the branes). In addition, we have the $D5$ branes  stretching to infinity which have non-normalizable open string modes turned on. 
There is a continuum of modes in the radial direction, but one needs only to explore the zero momentum modes in both the open and closed cases\foot{One can also compute the backreaction by integrating over the continuum, this gives the same answer as picking up the one-point function of the zero-momentum mode \murtro .}. 
We need to find out if the asymptotic profile of any of these modes changes, and whether any new non-normalizable modes get turned on. 

For our case $k=1$, this question turns out to have a further subtlety --  the tachyon winding mode turns  out to be at the edge of the normalizability bound, and one needs to be more careful. 
To understand this better, we shall make a small digression into the nature of vertex operators in this system.

\subsec{The normalizability of the vertex operators}

In general, systems with a linear dilaton have two branches of operators, one of which is normalizable in the radial direction and the other not. In the non-critical string, it was understood \SeibergEB\ that the local vertex operators were those that were non-normalizable. These were the operators that had a non-zero influence in the asymptotic region and defined the theory. Changing them is equivalent to changing the theory. 

This is consistent with our general notion of perturbative quantum gravity as well as ideas from holography where the bulk modes reaching the boundary would couple to operators which change the Lagrangian of the holographic theory. Taking into account the curvature of the space, the bound for an operator with the profile $e^{- j {\rho \over \sqrt{2k}}}$ is $Re(j) < \half$. The modes with $Re(j)=\half$ are delta function normalizable and are the travelling waves.

\vskip 0.5 cm

\ndt {\it Subtleties at $k=1$}

\vskip 0.2 cm

\ndt At $k=1$, the wavefunction for the tachyon winding mode is\foot{We have denoted $\rho + \t \rho$ by $\rho$ in this expression.}:
\eqn\wavefntach{\eqalign{
\phi^{tach}_{P}(\rho) 
& =  \cosh {\rho \over \sqrt{2}}  \; e^{\pm i \sqrt{1 \over 2}(\theta -\t \theta)} \; (\sinh{ {\rho \over \sqrt{ 2}}})^{-2iP-2} F\left(1+iP, 1+iP; 1+2iP;-{1\over \sinh^{2}{ {\rho \over \sqrt{ 2}}}}\right) +  \cr
& \qquad  R^{tach}(-P) \cosh  {\rho \over \sqrt{ 2}} e^{\mp i \sqrt{1 \over 2}(\theta -\t \theta)}  (\sinh{ {\rho \over \sqrt{ 2}}})^{2iP-2} 
 F\left(1-iP, 1-iP; 1-2iP;-{1\over \sinh^{2}{ {\rho \over \sqrt{ 2}}}}\right)   \,.
}}
where the reflection amplitude is (with $k=1$):
\eqn\reftach{
R^{tach}(P) = \nu^{2iP} { \Gamma(2iP)   \Gamma(1+2iP)  \Gamma(1-iP) \Gamma(-iP ) 
 \over  
\Gamma(-2iP)  \Gamma(1-2iP)  \Gamma(1+iP) \Gamma(iP) }. 
}
We can now see easily that the asymptotic behavior is
$\Phi^{tach}(\rho) \equiv T(\rho) = \mu e^{\pm i {1 \over \sqrt{2}} (\theta
-\t \theta) -  {\rho + \T \rho \over \sqrt{ 2k}}} $. This behavior is on the edge of the Seiberg window of
operators non-normalizable at the weak coupling end. We recall here a discussion from \Ashok\ which generalized the arguments of  \McGreevyEP\ 
to the supersymmetric theory.

First, we note that objects in the theory are singular in the limit $k \to 1$,
 and we regularize as $k = 1 -  \epsilon$. In order to keep quantities like
 the two and three point functions in the bulk theory finite, we need to keep
 $\mu { \Gamma(1/k) \over \Gamma(1-1/k)} $ finite \hoso . 
 This means that the bare $\CN=2$
 Liouville interaction  diverges.  To understand this, we shall follow the analysis in the $c=1$ theory \McGreevyEP\  and  look at
 the full wavefunction of the tachyon winding mode including the
 reflected piece. The reflection amplitude for the mode 
in the action $j = {k \over 2}$ has the value $R=-1$. The asymptotic behavior is (to leading order in $\epsilon$): 
\eqn\tachaftref{\eqalign{
\Phi^{tach}(\rho) \equiv T^{phys}(\rho) & = \mu \; \lim_{\epsilon \to 0} e^{\pm i \sqrt{1+\epsilon \over 2}(\theta -\t \theta)}  \left( e^{-(\sqrt{1-\epsilon \over 2})\rho} -  e^{-(\sqrt{1+\epsilon \over 2})\rho} \right) \cr 
& = (\sqrt{2} \mu \epsilon)\, \rho \, e^{\pm i {1 \over \sqrt{2}}(\theta -\t \theta)}  e^{- {1 \over \sqrt{2}} \rho} \equiv  \mu_{ren} \, \rho \, e^{\pm i {1 \over \sqrt{2}} (\theta -\t \theta) - {1 \over \sqrt{2}}\rho}  \cr
}}
where we have defined a quantity $\mu_{ren}$ which we should keep finite. 
Keeping track of all the terms in the above computation tells us that the full tachyon winding mode\foot{As was noted in \AharonyXN , it is difficult to say in the case $k=1$ whether there is an independent normalizable mode or not. We shall see that the D-branes shall backreact onto the non-normalizable modes and so our conclusions will be valid irrespective of the existence of such a mode.} 
 has also a normalizable subleading piece which behaves as $\Phi \sim \mu_{ren} \log \mu_{ren} e^{\pm i {1 \over \sqrt{2}}(\theta -\t \theta)-{1 \over \sqrt{2}} \rho}$.

\subsec{Open string backreaction}

As mentioned in section 2, the extended branes in this theory are semiclassically defined by a operator on the boundary worldsheet action multiplied by the so-called boundary cosmological constant $\mu_{B}$ [see e.g. \hoso\ for a discussion]. 
One can make arguments similar to the closed string case, about the operator  at $k=1$ -- there are two solutions, one of which behaves as $\rho e^{-{1 \over \sqrt{2}}\rho}$ and the other which behaves as $e^{-{1 \over \sqrt{2}}\rho}$. The non-normalizable mode multiplying $\mu_{B}$ is a linear combination of the above two. 

However, as discussed in section 2, the open string case is a little different from the closed string one in that there is a {\it genuinely} localized mode (bound state) in the spectrum. Using the understanding in the related $c=1$ theory \Teschner , it was argued \murtro\ that there must be exactly one localized scalar (superfield) fluctuation arising on the $\ket{D5,J=0}$ with the quantum numbers of the meson superfield. 
A further careful calculation \murtro\ of the spectrum on the $\ket{D5}$ brane using the methods of \TeschnerMD\ then bore this out.

Based on the above discussion, turning on a source for the meson on the magnetic brane would involve turning on the non-normalizable counterpart\foot{Note that this operator is the bottom component of the superfield whose top component is the actual six dimensional field which is the mass parameter in the electric theory.}  of the meson operator, {i.e.} the operator which decays as $\rho e^{-{1 \over \sqrt{2}} \rho}$.

$\mu_{B}$ is related to $(J,M)$ in our exact description  through a formula \hoso\ $\mu_{B} \bar{\mu}_{B}/\mu = (2k/\pi)\sin{\pi(J+M)} \sin{\pi(J-M)} $. For all the supersymmetric branes that we presented, if we pick $\mu$ and $\mu_{B} \bar{\mu}_{B}$ to be real and positive, $\mu_{B} \bar{\mu}_{B}$ is also real and positive. For the electric branes $M=\half, J=\half + iP$,  changing the real parameter $\mu_{B}$ moves the end of the brane up or down the cigar. In the magnetic branch $M=0, 0 \le J \le \half$, changing the real $\mu_{B}$ moves the localization of the magnetic field a little bit near the tip. In both cases, it changes the mass of the quarks.

It is natural to guess that the deformation $\a$ \deformchar\  in the semiclassical theory corresponds to a change in the value of $\mu_{B}$ in an imaginary direction (at least near $J=M=\a=0$), this is supported by the intuition that $\a$ rotates the branes around the sine-Liouville cylinder. This is a non-normalizable deformation; but work remains to be done to understand the semiclassics directly better for the branes outside the supersymmetric electric branch. 
We now turn to measure the  closed string backreaction  from the exact boundary state directly. 

\subsec{Closed  string backreaction}

Before computing the backreaction of the deformed boundary state, let us recall some facts about the backreaction of the supersymmetric D-branes. The calculation of the backreaction of say the $\ket{D3}$ brane onto the tachyon winding mode \Ashok\ was done by convolving the one-point function with the eigenmodes of the Laplacian on the cigar. This gave a result that the tachyon winding mode 
changed as $\delta T(\rho) =  {1 \over \epsilon} \Psi^{tach}(j=\half) \, e^{{1 \over \sqrt{2}}(\theta - \T \theta)  -{1 \over \sqrt{2}} \rho}, \, \rho \to \infty$ where $\Psi^{tach}(j=\half + ip)$ was the (finite) one-point function of the tachyon. There was a similar result for the backreaction onto the RR field at infinity which is the superpartner of the tachyon winding mode. 

The interpretation of this divergence was the following -- at infinity, the 
charge of the $\IR^{3,1}$ filling brane must be measured simply by  the field strength of the axion 
$\oint \p_{\theta} \chi$.
The normalization of the backreaction is such that one $\ket{D3}$ brane creates one unit of magnetic charge for the axion. 
The electric flux of this brane which is Hodge dual to the axion field strength is also constant and  the gauge field itself would asymptotically grow as $\rho$.  Upon action by the supercharge, the (RR) vertex operator\foot{See  \Ashok\ for a reasonably detailed discussion of the vertex operators in the various pictures in this context.} $g_{s} A_{0123} (\rho)$ produces the tachyon with a profile $\rho e^{-{1 \over \sqrt{2}}\rho}$.

The summary of the calculation is that the supersymmetric $D3$ branes  and $D5$ branes with $(J=M)$ backreact onto the non-normalizable tachyon winding mode and equally onto the $RR$ axion with values $1$  and $(\half -2M)$ respectively. 
Any change in the one-point function onto these modes will translate into a change of the asymptotic values of these fields. 
Since the branes are made up of only continuous representations, no other fields will get switched on, and the above two are the only fields we need to worry about.

To measure the backreaction, we simply need to pick out the one-point function of these two fields at $j=\half$ from the boundary state  \newcontbrane . 
The complex tachyon $T_{\pm}$ and the RR axion (also a complex scalar $\p_{\pm} \chi$ in four dimensions) both have $m=\half$. 
The one point functions of both these modes with $j=m=0$ do not depend on the value of $J$. 
The $\a$ dependence of the one-point functions comes from the overlap  \newishi\ between $\bbra{j=m=\half,0}q^{L_{0}} \kket{j=m=\half,\a }$. All four of the modes we are interested in have no bosonic oscillator excitation, and so have an overall factor of ${\sin(\pi \a) \over \cos(\pi \a) \sin(\pi \a)} = {1 \over \cos(\pi \a)}$. The tachyon winding mode $e^{\varphi \pm i k \theta + (|k|-{Q \over 2})\rho}$  which arises from the NS sector ground state of the fermionic oscillators is further multiplied by $1$. In the R sector, the ground states of the fermionic oscillators which gives rise to the axion is represented by the term $\half q^{1 \over 8} (e^{\pi i \a} + e^{- \pi i \a})$. On multiplying with the overall common factor and the normalizations of the closed string modes, the backreaction onto the tachyon is  $ \half {1 \over \cos(\pi \a)} = \half + { \pi^{2} \over 4} \a^{2} + O(\a^{4})$, whereas the backreaction onto the axion is $\half {\cos(\pi \a) \over \cos(\pi \a)}=\half$. 

\vskip 0.2cm

We can now compare the asymptotics of the supersymmetric brane configuration with a mass term in the electric theory (fig 5) and 
the new stable non-supersymmetric brane configuration at the end of the snapping process after introducing the mass term in the magnetic theory (fig 8). The semiclassics of these two configurations on the cigar were discussed in section 3, and the exact boundary states are  $N_{c}$ $\ket{D3}$  and $N_{f}$ $ \ket{J=M=\half + i \a}$ for the supersymmetric configuration  as opposed to $N_{c}$ $ \ket{J=M=0,\a}$ and $(N_{f} - N_{c})$ $\ket{J=\half + i  \a , M=\half,\a=0}$ for the stable non-supersymmetric configuration. 

As just computed above, the axion charge of the new stable non-supersymmetric branes is independent of $\a$, and thus the snapping process is thus allowed by charge conservation. On the other hand, we also see that the non-supersymmetric  configuration has different asymptotics than the supersymmetric one through the backreaction onto the tachyon winding modes.

\vskip 0.5cm



\centerline{\bf Acknowledgments}
I would like to thank Simeon Hellerman, Shamit Kachru, Soo-Jong Rey, Nathan Seiberg, David Shih and the members of the high energy group at the ASICTP and SISSA, Trieste for interesting discussions and questions. I would like to thank Jan Troost for many interesting conversations on these and related issues. 
I would especially like to thank Ofer Aharony for many interesting discussions, correspondence and comments on a preliminary draft. I would like to acknowledge the Institute for Advanced Study, Princeton and its members for hospitality while part of this work was being carried out.

\eject

\appendix{A}{Characters and modular transformations of the $\CN=2$ algebra}

The various functions used are:
\eqn\defs{\eqalign{
 \eta(\tau)& =e^{2 \pi i \tau \over 24} \prod_{n=1}^\infty (1-e^{2 \pi i n \tau}) \cr
 \vartheta_{ab}(\tau,z)& = \sum_{n\in {\bf Z}} \exp \left[ 2 \pi i \left(\half(n+{a \over 2})^{2} \tau + (n + {a \over 2})(z+{b \over 2}) \right)\right] \cr 
\Theta_{m,k} (\tau,z)& = \sum_{n \in \bf{Z}} \exp \left[ {2 \pi i \tau k (n+{m \over 2 k})^{2} + 2 \pi i z k (n+{m \over 2 k})}\right] = \vartheta_{{m \over k}0 } (2k\tau,kz) \cr
}}

The unflowed $(s=0)$ characters of the various representations are: 
\eqn\chars{\eqalign{
{\rm Continuous} \; (h,Q): & \quad  \chi^{cont}_{j,m} (\tau, z)  = q^{h-{1 \over 4k}} y^{Q} {\vartheta_{00} (\tau , z)  \over \eta^{3}(\tau)} \cr
{\rm Discrete} \; (h= \pm {Q \over 2}) : & 
 \quad \chi^{Disc}_{m} (\tau, z)  = {\chi_{j= \pm m}^{cont} (\tau, z) \over (1 + y^{\pm 1} q^{\half}) }\cr
{\rm Identity} \; (h=Q=0):  & \quad \chi^{Id} (\tau, z)  = {\chi^{cont}_{j=m=0} (\tau, z) (1-q) \over (1 + y^{-1}q^{\half}) (1 + y q^{\half})}  \cr
}}
The effect of spectral flow is (for any character)\foot{When the $s$ character label is not indicated, it means we set it to zero.}:
\eqn\specfl{
\chi_{*,s}(\tau,z) = q^{\half (1+{2 \over k}) s^{2}} y ^{(1+{2 \over k}) s} \chi_{*} (\tau,z + s \tau)
}
To have integral $U(1)_{R}$ charges after modular transformation, we need to consider the characters of the extended $\CN=2$ algebra, which include the spectral flow generators:
\eqn\exchardef{
Ch_{j,m,r}(\tau,z) = \sum_{s \in r + k \IZ} \chi_{j,m,s}(\tau,z)
}
The extended continuous characters can be expressed as:
\eqn\extcont{\eqalign{
Ch^{cont}_{j,m,r}(\tau,z)   & = q^{-{j(j-1) \over k}-{1\over 4}} \sum_{s \in r+ k \IZ } q^{{1 \over k}(s+m)^{2}} y^{{2\over k}(s+m)}   q^{{s^{2} \over 2}} y^{s} {\vartheta_{00} (\tau,z+s \tau)\over \eta^{3}(\tau)} \cr
& = q^{-{j(j-1) \over k}-{1\over 4}} \sum_{n \in  \IZ } q^{{1 \over k}(kn+r+m)^{2}} y^{{2\over k}(kn+r+m)}  {\vartheta_{2(kn+r),0} (\tau,z)\over \eta^{3}(\tau)} \cr
&= q^{-{j(j-1) \over k} -{1\over 4}} \Theta_{2(m+r),k}(\tau,{2z \over k}) {\vartheta_{2r, 0} (\tau,z)\over \eta^{3}(\tau)}  \cr
}}
They obey the modular transformation law:
\eqn\modtrans{
Ch^{cont}_{j,m,r}(\tau,z) =    \sum_{m' \in \IZ_{k}} e^{4 \pi i {m m' \over k}}  \int_{\half+i\IR^{+}} dj'  \cos{\left(4 \pi i(j-\half)(j'-\half) \right)}  Ch^{cont}_{j',m',0} ({-1 \over \tau}, {z \over \tau} + r)
}
Note above the periodicity of $m$ with period $k/2$. 
Note also that for $r \in \IZ$, the  function $\vartheta_{00}(\tau,z+r)$ is independent of $r$, and the 
expression above depends on $r$ only through the factor $e^{2 \pi i (m+r) {2m' \over k}}$. 
We thus keep $r \in [0,1)$.\foot{ For the boundary states that follow, different values of $r$ will correspond to gluing conditions that preserve different $\CN=1$ algebras on the boundary of the worldsheet and we shall only talk about $r=0$.}
The modular transformations of the discrete characters is a little more complicated, and we present them below for $k=1$.

Now we shall focus on $d=4$ which corresponds to $k=1$. The various extended characters (with $r=0$) are:
\eqn\konechars{\eqalign{
 {{\rm \bf Continuous:}}  & \quad  m \in \{0,\half \} \cr
 Ch^{cont}_{j,m}(\tau,z)  & = q^{-j(j-1)-{1\over 4}} \sum_{n \in \IZ} q^{(n+m)^{2}} y^{2(n+m)}   {\vartheta_{00} (\tau,z)\over \eta^{3}(\tau)} \cr
&= q^{-j(j-1)-{1\over 4}} \Theta_{2m,1}(\tau,2z) {\vartheta_{00} (\tau,z)\over \eta^{3}(\tau)}  \cr
 {\rm \bf Discrete:} &  \quad  j=|m| \in  \{0, \half \}  \cr
 Ch^{disc}_{m}(\tau,z) & = q^{-{1\over 4}} \sum_{n \in \IZ}{q^{n^{2}+2|m|n+|m|} y^{sgn(m)(2n+|m|)}    \over 
(1+y^{sgn(Q)} q^{n+\half})} {\vartheta_{00} (\tau,z)\over \eta^{3}(\tau)} \cr
{{\rm \bf Identity}}: &  \quad j=m=0 \cr
Ch_{Id}(\tau,z)  & = q^{-{1\over 4}} \sum_{n \in \IZ}{(1-q)q^{n^{2}+n-\half} y^{2n+1}    \over 
(1+y q^{n+\half})(1+yq^{n-\half})} {\vartheta_{00} (\tau,z)\over \eta^{3}(\tau)}. \cr
}}

The modular transformations for the $k=1$ characters are (with $\CS = \half + i \IR^{+}$):
\eqn\Strans{\eqalign{
Ch^{cont}_{j,m,\a}( \tau, z )   &= e^{i \pi {3 z^{2} \over \tau}} 2 \int_{\CS} dj' \cos{(4 \pi (j-\half)(j'-\half))} \times \cr
&  \qquad \left(   Ch^{cont}_{j',0}( -{1 \over \tau}, {z \over \tau}+\a) +  e^{- 2 \pi i m} Ch^{cont}_{j',\half}(  -{1 \over \tau}, {z \over \tau}+\a) \right). \cr
Ch^{disc}_{m=\pm1}(\tau, z ) & = e^{i \pi {3 z^{2} \over \tau}} \times \cr
& 
 \int_{\CS}  dj' \left[
\left( Ch^{cont}_{j',0}(\tau, z ) -  Ch^{cont}_{j', \half}(-{1 \over \tau}, {z \over \tau}) \right) \right.  - \left. {i \over 2} \left( Ch^{disc}_{m=\pm 1}(-{1 \over \tau}, {z \over \tau}) - Ch^{disc}_{m = \mp 1}( -{1 \over \tau}, {z \over \tau}) \right)  \right]. \cr
 Ch^{Id}(\tau, z)  & = e^{i \pi {3 z^{2} \over \tau}}  \times \cr
 & \int_{\CS} dj'   \left( - \sin^{2}{(\pi  (j'-\half) )} \; Ch^{cont}_{j',0}(-{1 \over \tau}, {z \over \tau} ) +  \cos^{2}{(\pi  (j'-\half) )} \; Ch^{cont}_{j',\half}(-{1 \over \tau}, {z \over \tau} ) \right). \cr
}}

The continuous extended characters of the $k=1$ theory obey the following two addition relations
\eqn\charaddn{\eqalign{
    Ch^{cont}_{j=m=\half}(\tau,z)  = &\; Ch^{disc}_{m=\half}(\tau,z) + Ch^{disc}_{m=-\half}(\tau,z)   \cr
    Ch^{cont}_{j=m=0}(\tau,z)  = & \; Ch^{cont}_{j=m=\half}(\tau,z)  +   Ch^{Id} (\tau,z). \cr
}}


\listrefs

\end